\documentclass[aip,graphicx, reprint]{revtex4-1}
\usepackage{braket}
\usepackage{tikz}
\usetikzlibrary{shapes.geometric, arrows}
\usetikzlibrary{graphs}
\usepackage[caption=false]{subfig}
\usepackage{color}
\usepackage{amsmath}
\usepackage{amssymb}
\usepackage{multirow}
\newcommand{\Argum}[1]{\ensuremath{\! \left( #1 \right)}}
\newcommand{\Tr}{\mbox{Tr}\;}
\newcommand{\TextSub}[1]{\ensuremath{_{\mbox{\tiny#1}}}}

\newcommand{\mc}[1]{\ensuremath{\mathcal{ #1 }}}  % My shorthand \mathcal{} command
\newcommand{\MyPar}[1]{\ensuremath{\left( #1 \right)}}
\usepackage{soul}

\begin{document}
\preprint{AIP/123-QED}
\title{An Explicit Electron-Vibron Model for Olfactory Inelastic Electron Transfer Spectroscopy}
 
\author{Nishattasnim Liza}
\author{Enrique P. Blair}
\email[]{enrique\_blair@baylor.edu}
\affiliation{Electrical and Computer Engineering Department, Baylor University, One Bear Place \#97356, Waco, Texas 76798, USA}
%%%%%%%%%%%%%%%%%%%%%%%%%%%%%%%%%%%%%%%%%%%%%%%%%%%%%%%%%%%%%%%%%%%%%%%%%%%%%%%%
\begin{abstract} 
The vibrational theory of olfaction was posited to explain subtle effects in the sense of smell inexplicable by models in which molecular structure alone determines an odorant’s smell. Amazingly, behavioral and neurophysiological evidence suggests that humans and some insects can be trained to distinguish isotopologue molecules which are related by the substitution of isotopes for certain atoms, such as a hydrogen-to-deuterium substitution. How is it possible to smell a neutron? The physics of olfaction may explain this isotopomer effect. Inelastic electron transfer spectroscopy (IETS) has been proposed as a candidate mechanism for such subtle olfactory effects: the vibrational spectrum of an appropriately-quantized odorant molecule may enhance a transfer rate in a discriminating electron transfer (ET) process. In contrast to other semi-classical or quantum-master-equation-based models of olfactory IETS, the model presented here explicitly treats the dynamics of a dominant odorant vibrational mode, which provides an indirect dissipative path from the electron to the thermal environment. A direct dissipative path to the environment also is included. Within this model, a calculation of ET rate is developed, along with a calculation of power dissipation to the thermal environment. Under very weak direct dissipative coupling, spectroscopic behaviors of the indirect path are revealed, and the resulting ET rate exhibits resonant peaks at certain odorant frequencies. Resonant peaks in ET rate also correlate to peaks in power dissipation. Spectroscopic behaviors are masked by strong direct dissipative coupling. Results support a rate-based discrimination between a preferred ligand and an isotopomer if indirect dissipative coupling dominates.
\end{abstract}
%%%%%%%%%%%%%%%%%%%%%%%%%%%%%%%%%%%%%%%%%%%%%%%%%%%%%%%%%%%%%%%%%%%%%%%%%%%%%%%%

\maketitle

%%%%%%%%%%%%%%%%%%%%%%%%%%%%%%%%%%%%%%%%%%%%%%%%%%%%%%%%%%%%%%%%%%%%%%%%%%%%%%%%
 
\section{Introduction}

Much insight  into the mechanisms of other senses has been developed, but mysteries persist in our modern understanding of olfaction. It is known that odorant molecules interact with olfactory receptors on sensory neurons, and that an odorant's structure is important in determining its scent, but there is evidence that additional, presently-unknown information may be required to distinguish molecules. For example, isotoptomers\textemdash molecules related by interchanging some atoms with isotopes\textemdash are nearly identical, yet some behavioral experiments suggest that humans and fruit flies can distinguish between isotopotomers by smell.\cite{2004KellerVosshall,2011FruitFly,2013-GaneTurin,Block2015,TurinPlausibilityLetter2015,2015BlockReply} 
Still, the dispute remains unresolved: \textit{in vitro} work also performed in selected mammalian olfactory receptors has not demonstrated a distinguishable response to isotopomer molecules,\cite{Block2015} but neurophysiological research has shown that insect antennae can produce a differential response to some isotopomers.\cite{2016HoneybeeAL,2016DrosophilaAntenna}

Olfactory models may be divided into two broad categories. One category may be termed ``lock-and-key'' models, or ``odotope'' models, in which odorant structure alone is assumed to determine the receptor response, just as a key's shape determines whether it can actuate a lock. A second class of models known as ``swipe-card'' models extend the lock-and-key model in that structure is recognized as necessary but insufficient to distinguish odors.\cite{2012QOlfactionReview} One prominent swipe-card theory invokes quantum mechanics: a suitable structure \textit{and} additional information encoded in the molecule trigger the olfactory response. This is in analogy to a magnetic hotel key card, which must not only fit in the swipe-card slot, but also must have the proper keyword encoded in the magnetic strip. 

\begin{figure*}[phtb]
  \includegraphics[width=\textwidth,height=7.2cm]{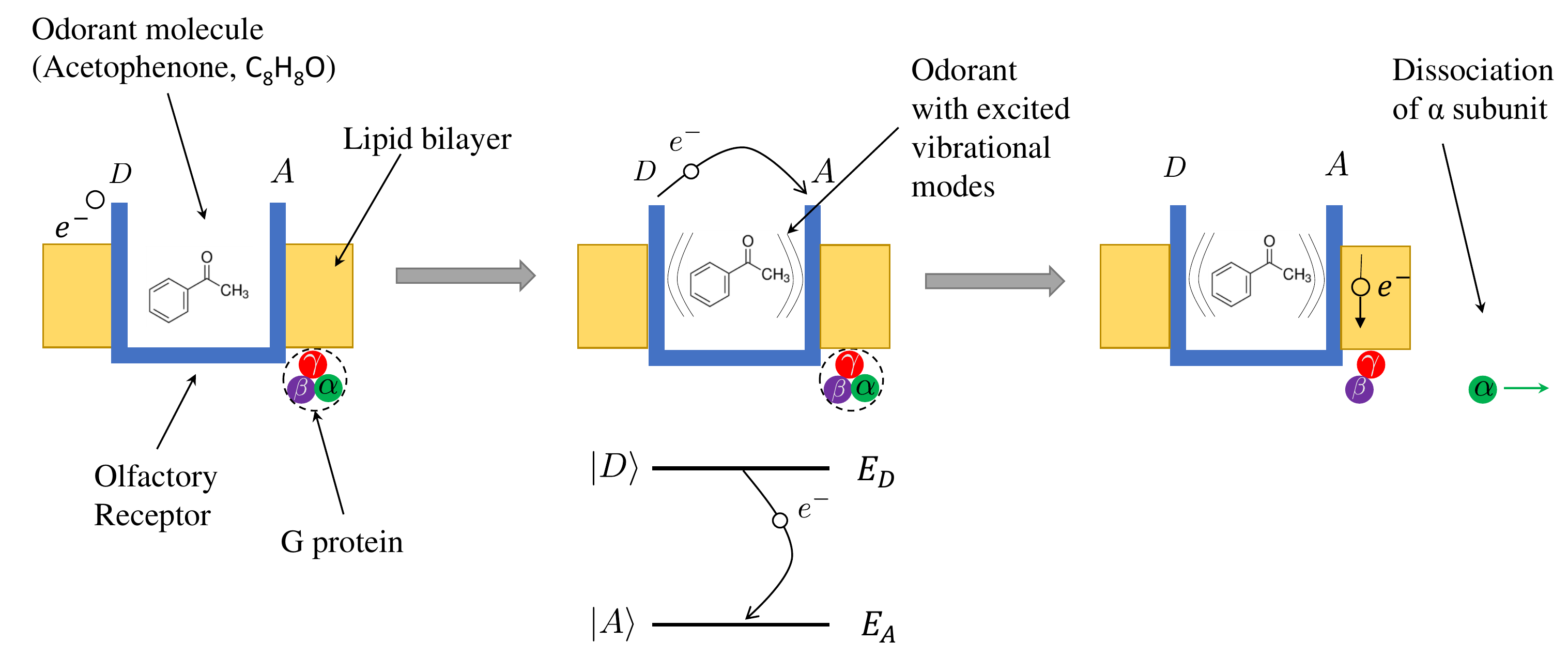}
  \caption{The inelastic electron tunneling spectroscopy (IETS) mechanism is illustrated in the context of human olfaction. A human odorant receptor is known to be a G-protein coupled receptor (GPCR, shown in blue, with an associated G protein). When an odorant molecule docks in the receptor (left figure), an electron transition from D (quantum state $\ket{D}$ at energy $E_D$) to A (state $\ket{A}$ at energy $E_A$) requires a change in energy that is facilitated by the excitation of a quantum of molecular vibrational energy. Once complete, the transferred electron may trigger the activation of the associated G protein, leading to an action potential and a sensory detection event. Without an odorant having a properly-quantized vibrational spectrum, the electron transition and subsequent detection are much less probable in the absence of other mechanisms by which the electron can relax from $E_D$ to $E_A$.}
  \label{fig:IETS_sketch}
\end{figure*}

Turin posited that spectroscopic information about a molecule could enhance an electron transfer (ET) rate, which would play an integral role in distinguishing between odorants.\cite{Turin1996,2018StatusVTO} After entering the nasal cavity and diffusing through the mucus layer, an odorant molecule may dock with an odorant receptor, a large protein spanning the bilipid cellular membrane. Here, the receptor is prepared with an electron on an donor site $D$ (electronic state $\ket{D}$, at an energy $E_D$). This is illustrated in the left-most portion of Figure \ref{fig:IETS_sketch}. Detection involves ET to an acceptor site $A$, at energy $E_A$, which is $\Delta$ below $E_D$. A direct $\ket{D} \rightarrow \ket{A}$ transition does not conserve energy, and thus is unlikely. On the other hand, if an odorant vibrational mode has quantization $\hbar \omega_o = \Delta$, then an inelastic $\ket{D} \rightarrow \ket{A}$ ET may readily occur with the associated excitation of a quantum of vibration in the odorant. A frequency-selective response of this type is referred to as inelastic electron tunneling spectroscopy (IETS), and an olfactory IETS process is depicted in the center part of Figure \ref{fig:IETS_sketch}. Then, the transferred electron is presumed to trigger some subsequent process, which continues the chain of events in sensing (in Figure \ref{fig:IETS_sketch}, this ET triggers the activation of the G protein, which releases an $\alpha$ subunit).
 
In this paper, we focus on the $\ket{D} \rightarrow \ket{A}$ transition without considering downstream and upstream events in the sequence leading to an olfactory detection. A fully-quantum, numerical model of the vibration-coupled IETS mechanism is presented here. 

Indeed, previous models of the vibrational theory of olfaction exist. Some of these models obtain an ET rate through Fermi's golden rule and are semi-classical in the sense that the electronic component is given a quantum treatment, and the vibrational components are treated classically. \cite{Brookes2007QuantumOlfaction,2012SolvyovPCCP} Other models go beyond Fermi's golden rule and obtain an ET rate from a master equation that includes the spectrum of the odorant and environmental harmonic oscillators.\cite{2012Bittner,2015Nazir,2017Tirandaz,2017-Brookes-Review}  In these models, the system is partitioned such that the degrees of freedom of the oscillators are traced over in the development of the master equation. In one of these models, the dissipation of energy (or power) was found to enhance a receptor's selectivity for a particular vibrational quantization in an odorant.\cite{2015Nazir} Another such model of chiral odorants showed that an energy difference between the ground states of enantiomer pairs could also lead to a rate-based dissipative discriminatory mechanism.\cite{2015chiralVTO}  While other models treat vibrational modes of the odorant and the environment as part of the reservoir and trace over these degrees of freedom to obtain a master equation, we develop a non-equilibrium model in which both the electron transfer and the dominant odorant vibrational mode are treated explicitly. This model lends itself to a calculation of power dissipation and highlights the link between power dissipation and electron transfer.

\section{Model}
\label{Model}

\subsection{Framework}
\label{Framework}

% ------------------------------------------------------------%
 \begin{figure}[bhpt]
 \centering
    \includegraphics[width=.95\columnwidth]{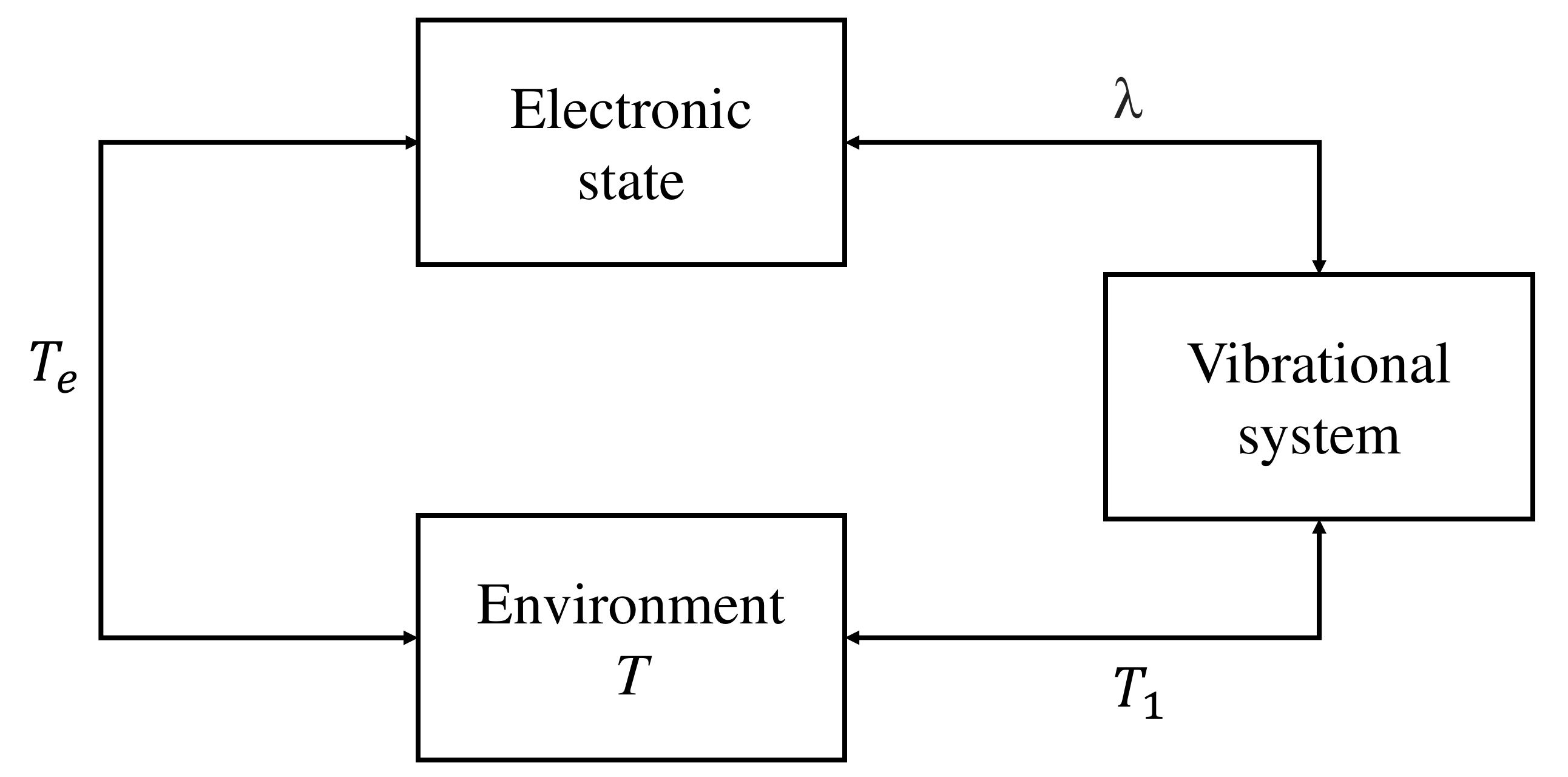}
\caption{The model developed in this paper includes the explicit, fully-quantum treatment of a $\ket{D} \rightarrow \ket{A}$ electron transfer event (ET) coupled to a dominant odorant vibrational mode. The ``Electronic state'' represents the ET event, and the ``Vibrational system'' represents the dominant odorant mode. This ET event is coupled to the thermal environment both directly, and indirectly via the odorant vibrational mode. In each case, the parameter that characterizes coupling between two systems is listed.}
\label{fig:ModelBlockDiagram}
\end{figure}
% ------------------------------------------------------------

The framework for this fully-quantum model of the vibrational theory of olfaction is an electronic two-state system dissipatively coupled in two ways to the thermal environment: (1) a direct coupling to environmental degrees of freedom, and (2) and indirect coupling via environmentally-damped odorant vibrations. The dynamics of the electronic system and one dominant odorant vibrational mode are explicitly treated in this paper. Dissipative effects are driven by interactions with the total environment, which is not explicitly modeled here but may include receptor vibrations and solvent degrees of freedom. The indirect dissipative path gives rise to spectroscopic behavior and is modeled using damped quantum oscillators to treat odorant vibrational modes. The direct electron-environment dissipation path is treated by building electronic relaxation into the model. This is depicted schematically in Figure\ \ref{fig:ModelBlockDiagram}. For computational tractability, only one odorant vibrational mode is presumed to be strongly dominant (the coupling between the electron and all other odorant modes is ignored). 
 
The electronic state is described by the Hamiltonian $\hat{H}_{e}= (\Delta/2) \hat{\sigma}_{z}-\gamma\hat{\sigma}_{x}$,
where $\hat{\sigma}_{x}$ and $\hat{\sigma}_{z}$ are Pauli operators, which we may write in terms of transition operators $\hat{P}_{jk}\equiv \ket{j}\bra{k}$ and projection operators $\hat{P}_{j} \equiv \ket{j} \bra{j}$: $\hat{\sigma}_{x} =\hat{P}_{AD} + \hat{P}_{DA}$ % $\ket{A}\bra{D}+\ket{D}\bra{A}$
and $\hat{\sigma}_{z} =\hat{P}_{D} - \hat{P}_{A}$.
%\hat{\sigma}_{z}=\ket{D}\bra{D}-\ket{A}\bra{A}$
The parameter $\gamma$ is the hopping energy between the two states $\ket{D}$ and $\ket{A}$, also known as the coupling constant, $H_{AB}$, in quantum chemistry. The detuning between the two states, $\Delta$, functions as the driving force behind the electron transfer: $\Delta=E_{D}-E_{A}$.

The dominant odorant vibrational mode is modeled as a quantum harmonic oscillator, with Hamiltonian $\hat{H}_v$:
% vvv------------------------------------------------------------vvv
\begin{equation}
\hat{H}_{v}=\left(\hat{a}^{\dagger}\hat{a}+\frac{1}{2}\right)\hbar\omega_{o}=\frac{\hat{P}^{2}}{2m_{o}}+\frac{1}{2}m_{o}\omega_{o}^{2}\hat{Q}^{2}.
\label{eqn:HvQHO}
\end{equation}
% ^^^------------------------------------------------------------^^^
Here, $\omega_o$ is the frequency of the dominant oscillator mode; $\hbar$ is the reduced Planck constant; and $m_o$ is the dominant vibrational mode's effective mass. The position (coordinate) operator $\hat{Q}$ and the momentum operator $\hat{P}$ may be written in terms of $\hat{a}^{\dagger}$ and $\hat{a}$, the creation and annihilation operators, respectively:
% vvv------------------------------------------------------------vvv
\begin{equation}
\hat{Q}=\sqrt{\frac{\hbar}{2m_{o}\omega_{o}}}\left(\hat{a}^{\dagger}+\hat{a}\right)
\quad \mbox{and} \quad
\hat{P}=i\sqrt{\frac{m_{o}\omega_{o}\hbar}{2}}\left(\hat{a}^{\dagger}-\hat{a}\right).
\end{equation}
% ^^^------------------------------------------------------------^^^
 The right-hand side of Eqn.\ (\ref{eqn:HvQHO}) is written in terms of the kinetic energy ($\hat{P}^{2}/2m_{o}$) and the potential energy $\left(m_{o}\omega_{o}^{2}\hat{Q}^{2}/2\right)$ operators for this vibrational mode. 

The coupling between the odorant and the receptor's electronic state is described by a linear coupling term in the Hamiltonian,
% vvv------------------------------------------------------------vvv
\begin{equation}
\hat{H}_{ev}=\frac{g_{ev}}{2}\hat{\sigma}_{z}\hat{Q}.
\end{equation}
% ^^^------------------------------------------------------------^^^
The coupling constant $g_{ev}$ depends on $\lambda $, the reorganization energy of the odorant's dominant vibrational mode: $ g_{ev}=\sqrt{m_o \omega_o^{2}\lambda}$. Thus, the fully-quantum system is described by the total Hamiltonian, $\hat{H}$, given by
% vvv------------------------------------------------------------vvv
\begin{align}
\hat{H} & =\hat{H}_{e}+\hat{H}_{v}+\hat{H}_{ev} \nonumber \\ 
 & =\frac{\Delta}{2}\hat{\sigma}_{z}-\gamma\hat{\sigma}_{x}+\frac{\hat{P}^{2}}{2m_{o}}+\frac{1}{2}m_{o}\omega_{o}^{2}\hat{Q}^{2}+\frac{g_{ev}}{2}\hat{\sigma}_{z}\hat{Q}.
\label{eq:HFullyQuantum}
\end{align}
% ^^^------------------------------------------------------------^^^

The effects of the bath on the electron+odorant system may be modeled by treating it as an open quantum system in a Markovian environment. To do this, we use the Lindblad equation,\cite{Lindblad1976}
% vvv------------------------------------------------------------vvv
\begin{equation}
\frac{d}{dt}\hat{\rho}=-\frac{i}{\hbar}\left[\hat{H},\hat{\rho}\right]+ \mathfrak{D} \, .
\label{eq:LindbladEvolution}
\end{equation}
% ^^^------------------------------------------------------------^^^
The density operator, $\hat{\rho} \Argum{t}$, describes the time-dependent state of the electron+odorant system. The first term describes unitary dynamics and is equivalent to the quantum Liouville equation. The dissipator, $\mathfrak{D}$, models environmental effects and is given by
% vvv------------------------------------------------------------vvv
\begin{equation}
\mathfrak{D} = \sum_{j=1}^{s}\hat{L}_{j}\hat{\rho}\hat{L}_{j}^{\dagger}-\frac{1}{2}\left\{ \hat{L}_{j}^{\dagger}\hat{L}_{j},\hat{\rho}\right\} .
\label{eq:dissipator}
\end{equation}
% ^^^------------------------------------------------------------^^^
The Lindblad operators, $\{\hat{L}_j\}$, also known as environmental channels, describe the effects of the environment on the odorant-receptor complex. In this model, the indirect-path Lindblad operators are chosen so as to damp only the vibrational subsystem. In particular, two Lindblad operators will be used:
% vvv------------------------------------------------------------vvv
\begin{equation}
\hat{L}_1 = \frac{1}{\sqrt{T_1}} \hat{a}, \quad \mbox{and} \quad \hat{L}_2 = \exp \Argum{-\frac{\hbar \omega_o}{2 k_B T} } \frac{1}{\sqrt{T_1}} \hat{a}^{\dagger} .
\end{equation}
% ^^^------------------------------------------------------------^^^
The operator $\hat{L}_1$ removes energy from the vibrational system, with $T_1$ being a phenomenological characteristic time for exponential energy relaxation via this indirect path. After a time $t \gg T_1$, the system will relax to its ground state if only $\hat{L}_1$ is used. The operator $\hat{L}_2$ excites the system and includes a prefactor which depends on temperature $T$. When both $\hat{L}_1$ and $\hat{L}_2$ are used in concert, they drive the system to a Boltzmann distribution. Finally, the direct electron-environment coupling is modeled using an additional pair of Lindblad operators, $\hat{L}_3$ and $\hat{L}_4$, given by:
% vvv------------------------------------------------------------vvv
\begin{equation}
\hat{L}_3 = \frac{1}{\sqrt{T_e}} \hat{P}_{DA}, \quad \mbox{and} \quad \hat{L}_4 = \exp \Argum{-\frac{\Delta}{2 k_B T} } \frac{1}{\sqrt{T_e}} \hat{P}_{AD} .
\end{equation}
% ^^^------------------------------------------------------------^^^
 
Here, $T_e$ is the characteristic time for an exponential relaxation via the direct electron-environment dissipation path. The combination of $\hat{L}_3$ and $\hat{L}_4$ drives the system to a Boltzmann distribution.

To quantify the strength of enviromental coupling, it is helpful to define coupling strength ratios. For indirect coupling, the harmonic oscillator's period $T_o = 2\pi/\omega_o$ is the natural time scale, so $\chi_v \equiv T_o/T_1$ provides a useful measure of vibron-environment coupling. A vibrational system decoupled from the environment is characterized by $\chi_v \rightarrow 0$. For the electronic system, the detuning, $\Delta$, is the natural energy scale, so we use $\chi_e \equiv t_{\Delta}/T_e$ to characterize the strength of direct electron-environment coupling, where $t_{\Delta} \equiv \hbar/\Delta$. Small $\chi_e$ characterizes an electronic system for which direct environmental dissipation is suppressed. Additionally, when comparing the two coupling strengths, we use the ratio $\xi \equiv T_e/T_1$, for which $\xi \rightarrow 0$ when direct electron-environment coupling is dominant, and $\xi \rightarrow \infty$ for dominant indirect electron-environment coupling via the odorant vibrational mode.

In order to qualitatively show that in the regime of interest, the fully-quantum model is preferable over a semi-classical treatment, we reduce the fully-quantum model to a semi-classical treatment. If the kinetic energy of the oscillator is ignored, and if the coordinate is treated as a classical variable instead of as an operator ($\hat{Q}\rightarrow Q$), then the Hamiltonian reduces to a Marcus-type Hamiltonian, $\hat{H}^{\left(M\right)}$, which acts on the Hilbert space of the electronic system only:
% vvv------------------------------------------------------------vvv
\begin{equation}
\hat{H}^{\left(M\right)}=\frac{\Delta}{2}\hat{\sigma}_{z}-\gamma\hat{\sigma}_{x}+\frac{1}{2}g_{ev}\hat{\sigma}_{x}Q^{2}+\frac{1}{2}m_{o}\omega_{o}^{2}Q^{2}.
\end{equation}
% ^^^------------------------------------------------------------^^^

 % ------------------------------------------------------------ 
 \begin{figure}[hpbt]
 \begin{center}  
     \subfloat[$\lambda$ = 30 meV]{\includegraphics[width=1\columnwidth]{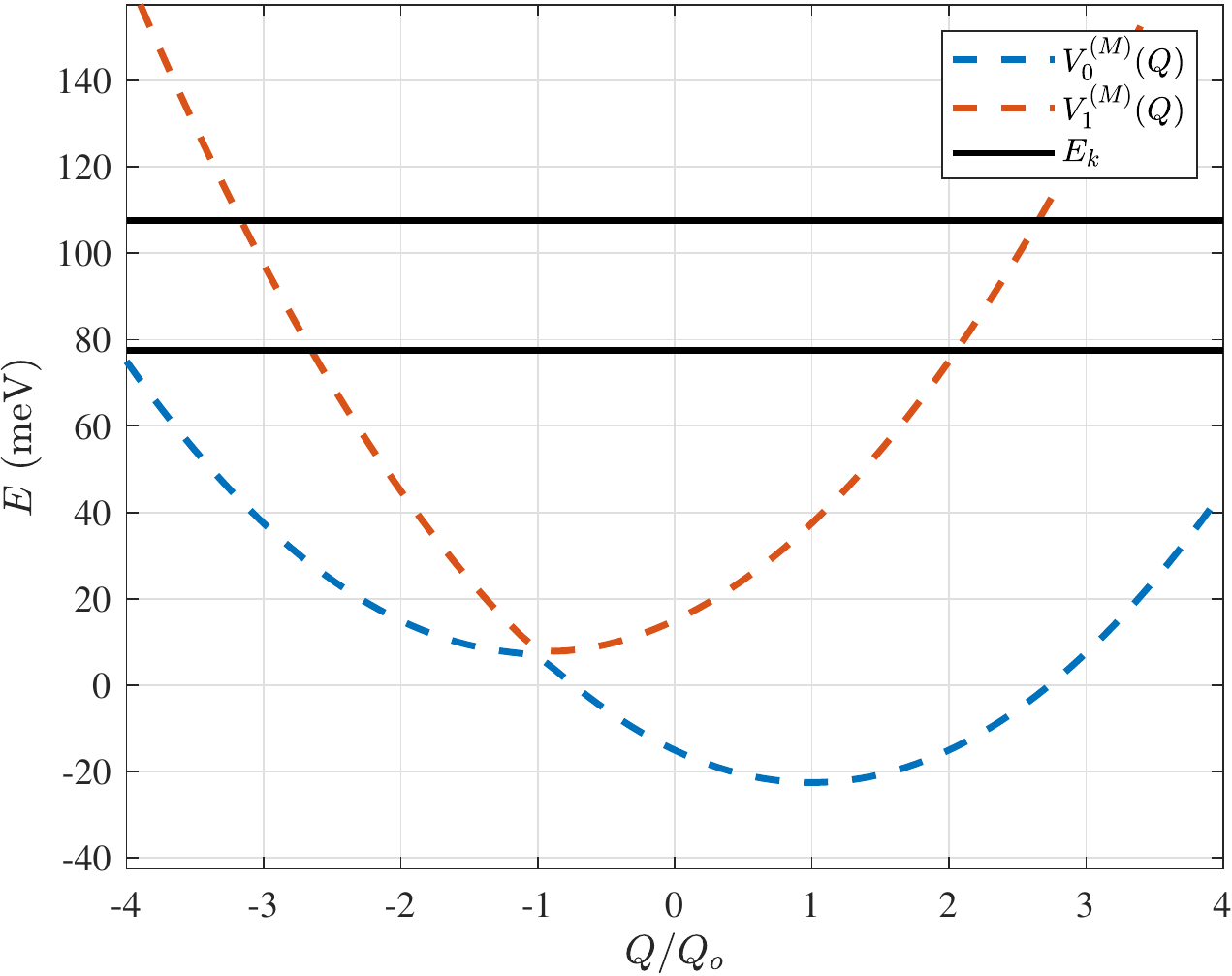}}
     
     \subfloat[$\lambda$ = 300 meV]{\includegraphics[width=1\columnwidth]{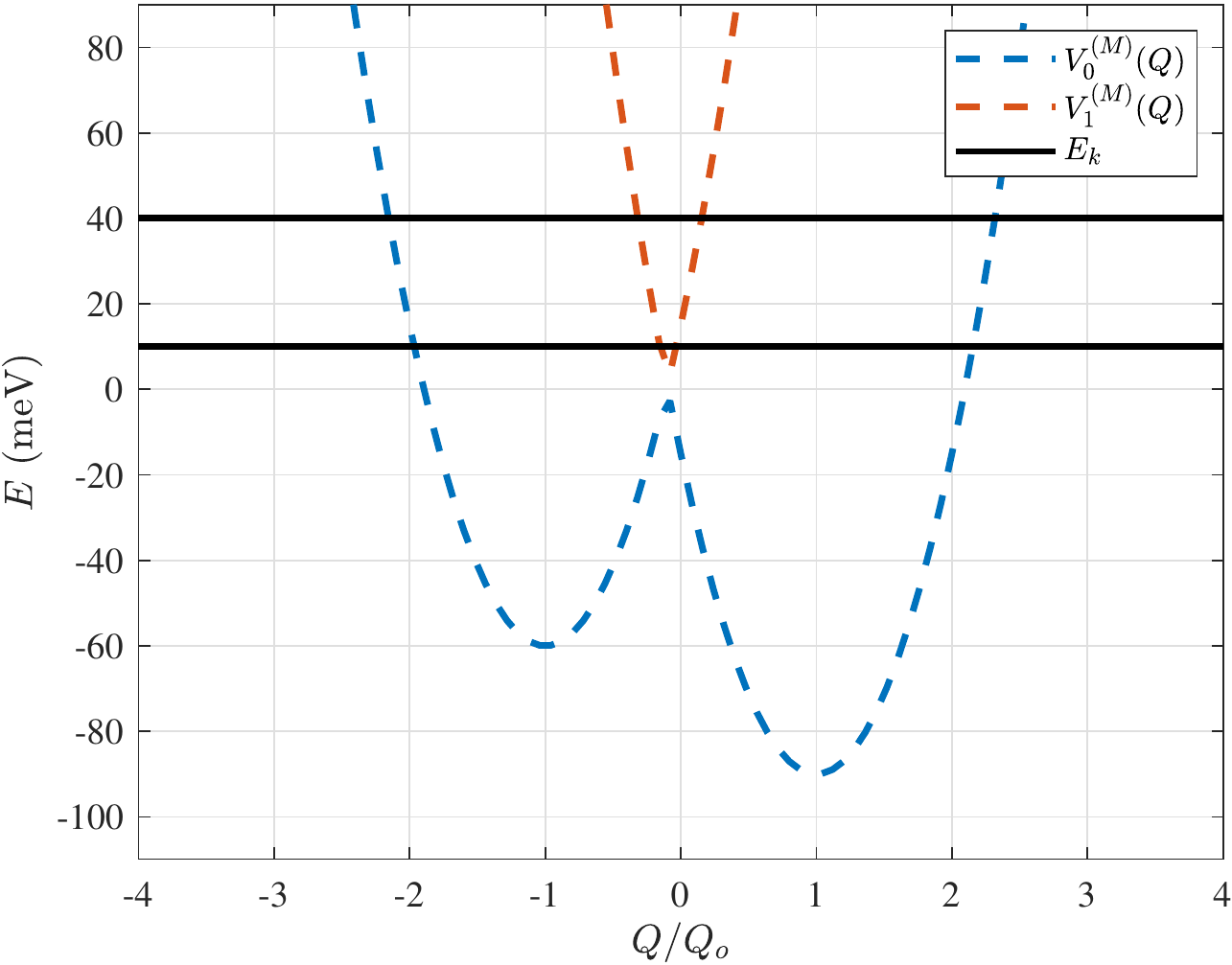}}
    \end{center} 
\caption{A fully-quantum treatment within the vibrational theory of olfaction is justified. The eigenvalues of the fully-quantum Hamiltonian $\hat{H}$ are plotted relative to the Marcus-type potential energy surfaces $V^{(M)}_k (Q)$ vs. $Q$. This plot is shown for two different values of the odorant reorganization energy $\lambda$: once with $\lambda=30~\mbox{meV}$ and again with $\lambda=300~\mbox{meV}$. In each case, the energy discretization of fully-quantum system is significant compared to feature sizes of semi-classical adiabatic Marcus-type potential surfaces. This suggests that treatment using the fully-quantum model will lead to behaviors not captured by a Marcus-type semi-classical model.}
   \label{fig:marcus12}
\end{figure}
% ------------------------------------------------------------ 

% ------------------------------------------------------------ 
\subsection{Choice of Parameters}

In the absence of concrete experimental data for model parameters, we pick typical values for biological systems and theorized values from the literature. 
Thus, all calculations are performed at ambient temperature ($T=293~\mbox{K}$) unless the role of temperature is investigated. A tunneling energy $\gamma = 1 ~\mbox{meV}$  
 is chosen, in following with analysis found in the literature.\cite{Brookes2007QuantumOlfaction}  $D$ and $A$ are treated as single molecular orbitals coupled to each other by a weak hopping integral $\gamma$, as there should be essentially no tunneling from $D$ to $A$ in the absence of the odorant or any other dissipative pathway. % Discussions in the literature use reorganization energy values in the range $30-300$ meV.
We often use a $D$-$A$ detuning of $\Delta \sim 200$ meV (1613.6 cm$^{-1}$), chosen since the interesting range in olfaction is 70 meV $< \hbar w_o <$400 meV ( 564.77 cm$^{-1}$ $< \hbar w_o <$ 3227.3 cm$^{-1}$).\cite{Brookes2007QuantumOlfaction, 2018OlfactoryBand} The odorant reogranization energy is varied between $30~\mbox{meV}$ ($\sim k_B T$ at biological temperatures) and $300~\mbox{meV}$ ($\sim \Delta$). 
 
% ------------------------------------------------------------ 
\begin{table}[ht!]
\centering
\begin{tabular}{||c|c ||}  
\hline
\textbf{Parameter} & \textbf{Value}  \\[0.5ex] 
\hline\hline
$\Delta$  & 200 meV ( 1613.6 cm$^{-1}$)\\
\hline
 $\gamma$ & 1 meV \\ 
 \hline
$\lambda$ & 30-300 meV \\ 
\hline
  $T$ & 293 K\\[1ex] 
 \hline 
\end{tabular}
\caption{Physically-interesting values of model parameters are chosen in following with treatments from the literature and are enumerated here.\cite{Brookes2007QuantumOlfaction, 2018OlfactoryBand}}
\label{table:2}
\end{table}
% ------------------------------------------------------------ 
% ------------------------------------------------------------ 
\subsection{Justification for a Fully-quantum Treatment}
\label{Justification}

A comparison of the statics between the fully-quantum treatment and the semiclassical Marcus-type reduction reveals the necessity for a fully-quantum treatment of IETS within this model. The stationary states $\ket{\phi_k}$ and eigenvalues $E_k$ for the Hamiltonian of Eqn.\ (\ref{eq:HFullyQuantum}) are found by solving the time-independent Schr\"odinger equation:
\begin{equation}
\hat{H} \ket{\phi_k}= E_k \ket{\phi_k}
\label{eq:justify1}
\end{equation}
The semiclassical $\hat{H}^{(M)}$ has two eigenstates, $\{\ket{\phi^{(M)}_k} \}$, with $k\in {1,2}$, and two eigenvalues $V^{(M)}_k$ , which solve the time-independent Schr\"odinger equation:
 \begin{equation}
\hat{H}^{(M)}\Ket{\phi^{(M)}_k}= V^{(M)}_k \Ket{\phi^{(M)}_k}
\label{eq:justify2}
\end{equation}
Since $\hat{H}^{(M)}$ is a function of the coordinate $Q$, so also are the eigenvalues $V^{(M)}_k \Argum{Q}$ and eigenstates $\ket{\phi^{(M)}_k \Argum{Q}}$. The eigenvalues $V^{(M)}_k \Argum{Q}$ provide an adiabatic potential-energy landscape for the semi-classical system.
Figure \ref{fig:marcus12} shows the potential energy surfaces defined by the $\{ V^{(M)}_k \Argum{Q} \}$ for the semiclassical system, along with the lowest-energy eigenvalues for the fully-quantum system. The $\{V^{(M)}_k (Q) \}$ surfaces are plotted with $Q$ scaled to $Q_0 = g_{ev} /2m_o \omega^2_o$.  The energy quantization in the fully-quantum treatment is significant compared to the feature sizes seen in the potential landscape of the semi-classical treatment. This indicates that the fully-quantum treatment of the electron+odorant system is justified, indeed. Thus, the fully-quantum treatment will yield results which semi-classical models cannot capture. This comparison is performed for two different values of $\lambda$. It is seen that for a larger $\lambda$, the potential barrier between the coordinates $Q=\pm 1$ becomes more significant.

 \subsection{Calculation of electron transfer time, $t_{ET}$}
\label{Calc_ETrate}

% ------------------------------------------------------------
\begin{figure}[bthp]
      \begin{center}  
      \includegraphics[width=1\columnwidth]{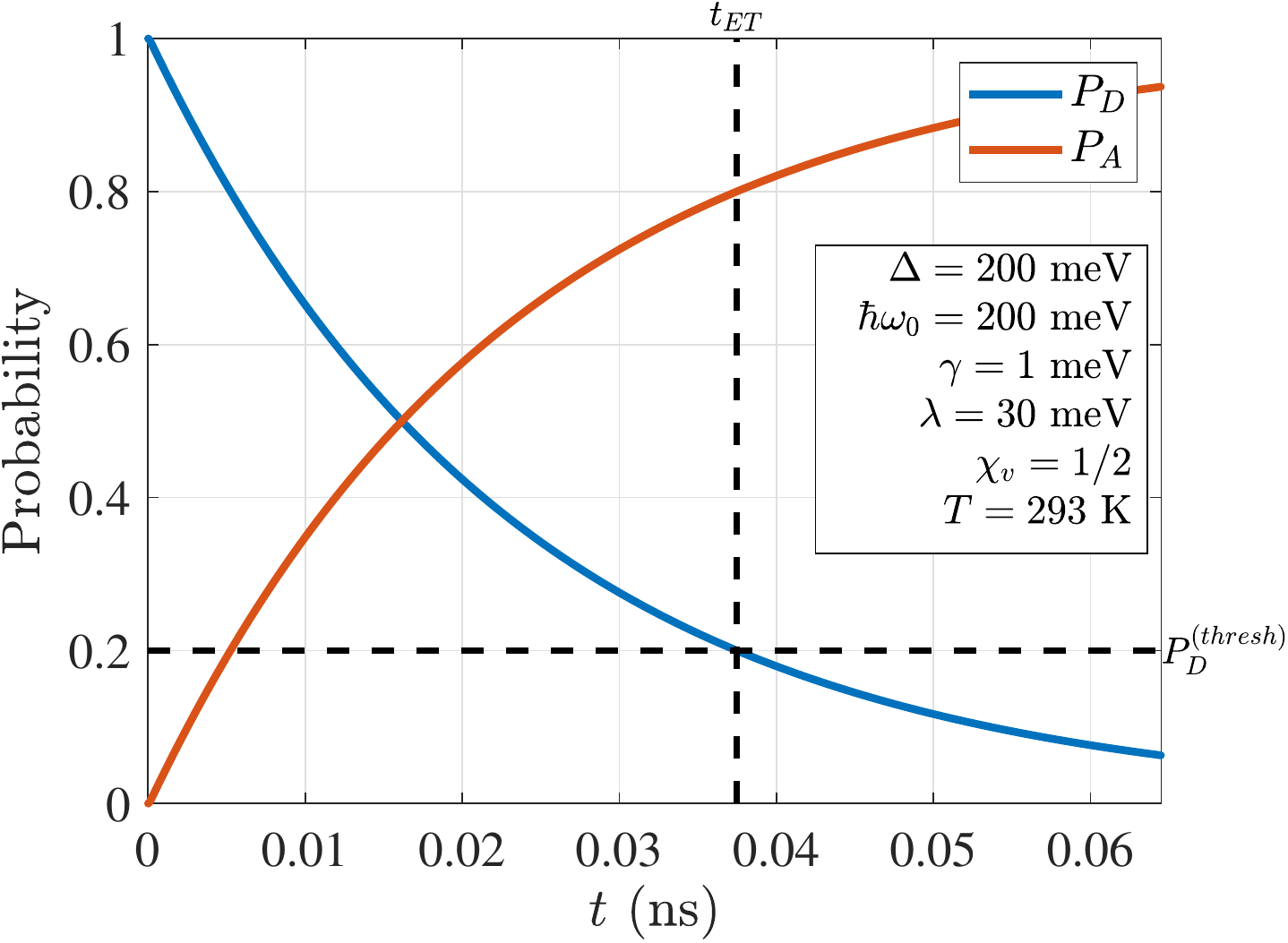}
      \end{center}  
      \caption{Calculation of threshold defined electron transition time, $t_{ET}$,  from the dynamic solution to  Eqn.\ (\ref{eq:LindbladEvolution}). Here, $\hat{\rho}(t)$ is calculated and the probabilities $P_D$ and $P_A$ for finding the electron on $D$ or $A$ are calculated. $t_{ET}$ is defined as the time when $P_D$ drops from $P_D(0)\simeq 1$ to a threshold value $P^{(thresh)}_D$. When $P^{(thresh)}_D = 0.2$, the result is $t_{ET}= 37.5~\mbox{ps}$. % 0.037473 ns.
Dashed black lines indicate the threshold $P^{(thresh)}_D = 0.2$ and the electron-transfer time, $t = t_{ET}$, defined as the time that $P_D$ drops below $P^{(thresh)}_D$.}
      \label{fig:t_ET}
   \end{figure}
% ------------------------------------------------------------
% ------------------------------------------------------------
Electron transfer times, $t_{ET}$, within the fully-quantum treatment are obtained from the time dynamics of the non-equilibrium model calculations. First, the system is prepared in an initial state $\hat{\rho}( t \leq 0)$, which is chosen to be, $\hat{\rho}^{th}_0$, the thermal equilibrium density matrix for the electron+ odorant system:
$$\hat{\rho}(0)=\hat{\rho}^{th}_0 \equiv \frac{1}{Z}\exp \Argum{-\frac{1}{k_B T} \hat{H}_0},$$with $$Z=\Tr \left(\exp \Argum{-\frac{1}{k_B T} \hat{H}_0}\right).$$
Here, $\hat{H}_0$ is the similar to $\hat{H}$ from
Equation\ (\ref{eq:HFullyQuantum}), but differs in that we set $\Delta \Argum{ t \leq 0} = \Delta_0$.  $\Delta_0$ is a large negative potential applied for the sole purpose of confining the electronic state to $\ket{D}$. Then, at $t = 0$, $\Delta \Argum{t}$ is changed abruptly to a positive, static value appropriate to the physics of the receptor. Next, $\hat{\rho}(t)$ is obtained for $t > 0$ by solving Eqn.\ (\ref{eq:LindbladEvolution}) numerically using the new, constant $\hat{H}$ for $t > 0$, and using $\hat{\rho}(0)$ as the initial value. The probabilities for finding the electron on the donor site and the acceptor sites, $P_D$ and $P_A$, respectively, are calculated as the expectation value of the projection operators: $P_D= \braket{\hat{P}_D}$ and $P_A= \braket{\hat{P}_A}$.

Figure\ \ref{fig:t_ET}  shows non-equilibrium model data from which an electron transfer time $t_{ET}$ is calculated. $t_{ET}$ is defined as the time when $P_D$ drops from $P_D(0)\simeq 1$ to a threshold value $P^{(thresh)}_D$. The electron transfer rate $k$ is the reciprocal of the electron transfer time: $k = 1/t_{ET}$. While any $P^{(thresh)}_D < 0.5$ could serve as a threshold, we choose $P^{(thresh)}_D = 0.2$ because this is low enough to preclude complicating oscillations in $P_D \Argum{t}$ as $P_D \Argum{t} \rightarrow P^{(thresh)}_D$, and yet high enough to allow reasonable calculation times. This calculation is performed with large $\xi$ so that indirect dissipation is dominant. Enhancing the direct dissipation pathway (reducing $\xi$) would allow a faster relaxation, resulting in a faster ET and a higher rate $k$.

The Rabi oscillation frequency $\gamma/\pi\hbar$ sets the upper speed limit for quantum charge transport in this system in the absence of coherent driving, so half of the Rabi oscillation period sets the lower limit for physical electron transfer times: $t_{ET} > \pi\hbar/2\gamma$. Here, with  $P^{(thresh)}_D = 0.2$, $t_{ET}= 37.5$ ps, which satisfies $t_{ET} > \pi\hbar/2\gamma \simeq 1~\mbox{ps}$.
The result for $t_{ET}$ is plausible because $t_{ET}$ does not violate its lower limit $\pi\hbar/2\gamma$ and the result is well below the biological time scale of ms for actuating GPCRs.\cite{GPCRTimescale}
  
 \subsection{Calculation of power dissipation, $\bar{p} \TextSub{diss}$}
\label{Calc_power_diss}

  Power dissipation may be calculated within this model. Power dissipation here is given by 
  %This calculation is summarized here, but a full derivation of power dissipation is given in Appendix \ref{appendix:power_flow}. 
% vvv--------------------------------------------------------------------------vvv
\begin{equation}
p \TextSub{diss} = -p_{\mc{E},ev} - p_{\mc{E},v}  - p_{\mc{E},e}\; , \label{eqn:dissipation}
\end{equation}
% ^^^--------------------------------------------------------------------------^^^
where
% vvv--------------------------------------------------------------------------vvv
\begin{align}
p_{\mc{E},ev} & = \Tr \MyPar{ \mathfrak{D} \hat{H}_{ev}} \, ,
\nonumber \\
p_{\mc{E},v} & = \Tr \MyPar{ \mathfrak{D} \hat{H}_{v}} \, , \, \mbox{and}
\nonumber \\
p_{\mc{E},e} & = \Tr \MyPar{ \mathfrak{D} \hat{H}_{e}} \, .\label{eqn:WorkDoneByEnv}
\end{align}
% ^^^--------------------------------------------------------------------------^^^
We interpret $p_{\mc{E},ev} + p_{\mc{E},v}+ p_{\mc{E},e}$ as the rate of work done on the electron+odorant system by the environment.

Average power dissipation, $\bar{p} \TextSub{diss}$, is evaluated over the electron transfer time $t_{ET}$:
% vvv--------------------------------------------------------------------------vvv
\begin{equation}
\bar{p} \TextSub{diss} = \frac{1}{t_{ET}} \int_0^{t_{ET}} p \TextSub{diss} \Argum{s} ds \; .
\label{eqn:PowerDissipation}
\end{equation}
% ^^^--------------------------------------------------------------------------^^^

% ------------------------------------------------------------

\section{Results}

Results are presented in three different regimes: (A) in the limit of dominant indirect coupling (weak direct electron-environment coupling, $\xi \rightarrow \infty$); (B) in the limit of dominant direct electron-environemtn coupling (weak indirect electron-phonon-environment coupling, $\xi \rightarrow 0$); and (C) an intermediate regime, in which both couplings are present, but neither is dominant.
\label{Results}

\subsection{Indirect Coupling Dominates}

When indirect coupling provides the dominant dissipation pathway, a rich set of spectroscopic behaviors is seen. Here, we explore the relationship between rate and odorant frequency, odorant reorganization energy, tunneling energy and temperature, and power dissipation.

\subsubsection{Resonant Peaks in ET Rate}
\label{res_peak}

\begin{figure}[bthp]
      \centering
      \includegraphics[width=1\columnwidth]{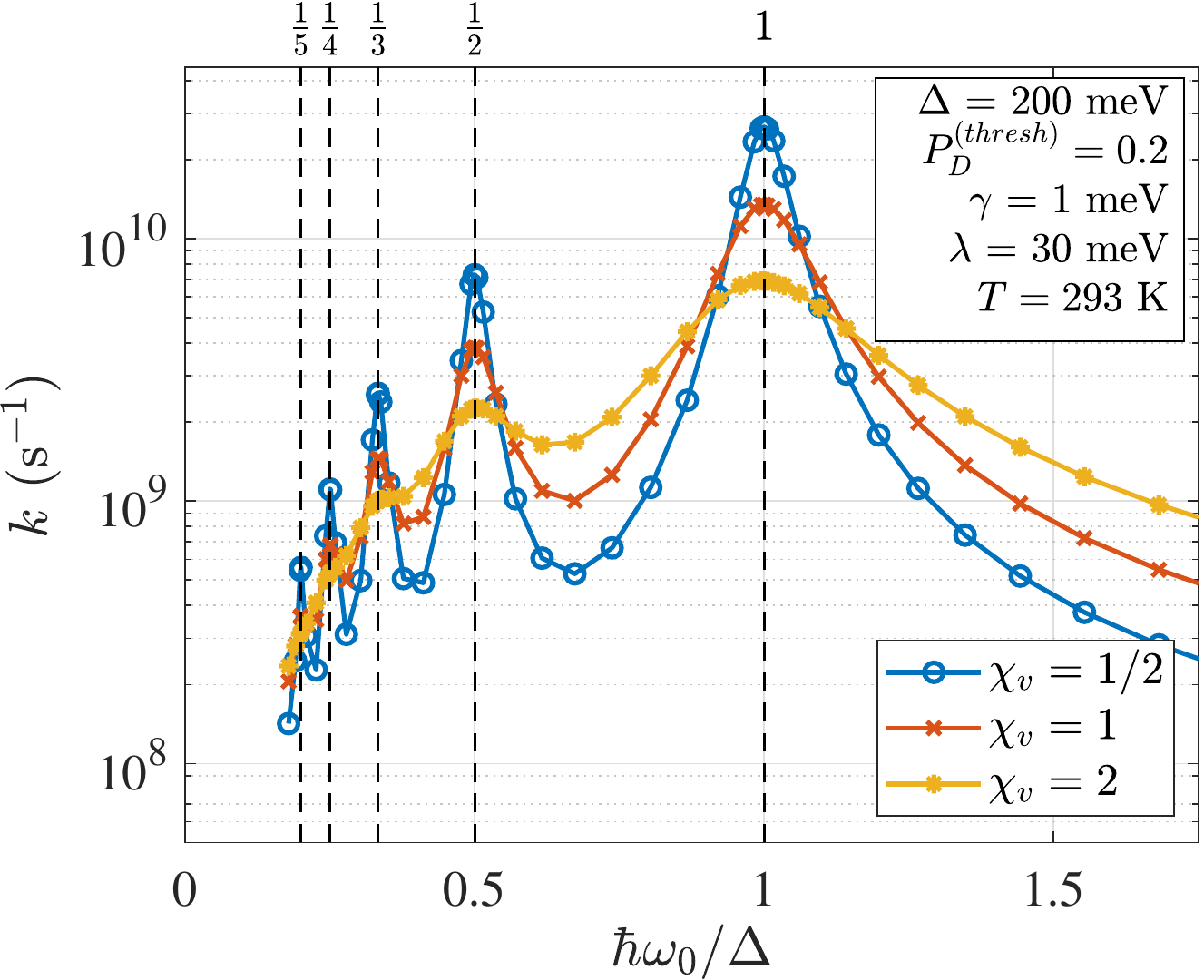}
      \caption{The electron transfer rate, $k$, from state $\ket{D}$ to state $\ket{A}$ exhibits resonant tunneling and environmentally-driven broadening of peaks. Here, peaks occur under a resonant condition that arises when the detuning $\Delta = E_D - E_A$ is equal to an integer multiple of the vibrational quantization $\hbar \omega_o$. Additionally, resonant peaks are broadened as odorant-environment coupling increases, as quantified by $\chi_v$.}
      \label{fig:compare_coupling}
   \end{figure}

When indirect coupling is dominant ($\xi \rightarrow \infty$), electron transfer rates exhibit resonant peaks. Figure\ \ref{fig:compare_coupling} shows the ET rate $k$ as a function of odorant vibrational angular frequency $\omega_o = 2\pi f_o$.  The frequency axes is scaled in units of $\Delta = E_D - E_A$. Resonant peaks in $k(f_o)$ occur at frequencies $\omega_o = \Delta / s \hbar$, where $s$ is a positive integer. As odorant environmental coupling ($\chi_v$) increases, the spectral peaks broaden.

% ------------------------------------------------------------
\begin{figure}[thpb]
      \centering
      \includegraphics[width=0.45\textwidth]{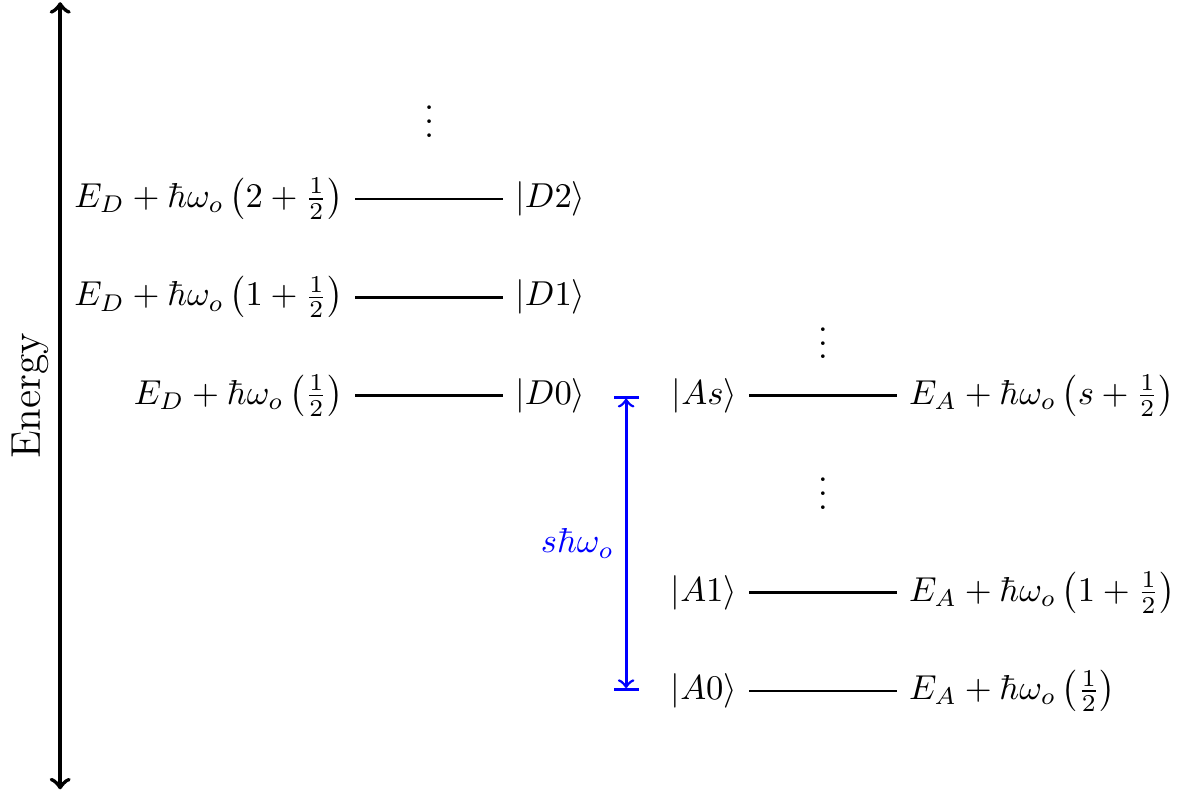}
      \caption{The eigenenergy spectrum for the system results in a resonant effect for the ET rate $k$. In the limit of $\gamma = \lambda = T = 0$, the eigenstates are $\ket{Xn}$, with energies $E_X + \hbar \omega \left(n + 1/2\right)$. When $\Delta = E_D - E_A = s\hbar \omega$ for some integer $s$, a resonant transition $\ket{D0} \rightarrow \ket{As}$ is favorable, as is the resonant back-transition$\ket{As} \rightarrow \ket{D0}$. The probability $P_D$ does not decrease to a small value until the system can dissipate energy for the state to approach $\ket{A0}$. Since this dissipation takes time, the minimum non-zero value ($n=1$) enables the fastest dissipation and thus the fastest ET time and the highest ET rate $k$.} 
      \label{fig:IETS_resonance_sketch}
   \end{figure}

% ------------------------------------------------------------
The resonance of Figure\ \ref{fig:compare_coupling} can be understood in the limit of $\lambda=0$, $\gamma = 0$ and $T=0$. Here, the eigenstates of the system are product states $\ket{X}\otimes \ket{n} = \ket{Xn}$, with eigenenergies $E_{X} + \hbar \omega_o \left( n + 1\right)/2$ for occupation number $n \in \left\{ 0, 1, 2, \ldots \right\}$ and $X \in \{D, A\}$. This spectrum is depicted in Figure\ \ref{fig:IETS_resonance_sketch}. The initial state is approximately $\ket{D0}$, with energy $E_D + \hbar \omega_o/2$. If $\Delta = E_D - E_A = s \hbar \omega_o$ for some positive integer $s$, then the initial energy $E_D + \hbar \omega_o/2$ matches exactly the eigenenergy $E_A + \hbar \omega_o \left( s + 1 \right)/2$, facilitating an $D\rightarrow A$ electron transfer event from the combined state $\ket{D0}$ to $\ket{As}$. In order for the system to settle to $\ket{A0}$ and for the electron transfer to ``complete'' ($P_D \rightarrow 0$), the energy $s \hbar \omega_o$ must be dissipated. Thus, low $s$ minimizes the vibrational energy dissipation and time required for the ET event, and peaks in the rate $k$ grow larger with decreasing $s$, with $s=1$ providing the highest peak. Multi-phonon processes\textemdash infrequently-considered in the theory of olfactory IETS\textemdash could play a role. In this model, ET rates are increased when the energy of fewer phonons must be dissipated to allow the system to settle to $\ket{A0}$.

% ------------------------------------------------------------
% ------------------------------------------------------------

\subsubsection{Odorant Reorganization energy, $\lambda$ }
 % ------------------------------------------------------------  
   \begin{figure}[hpbt]
\centering
\subfloat[$\; \chi_v=1/2$]{\includegraphics[width=1\columnwidth]{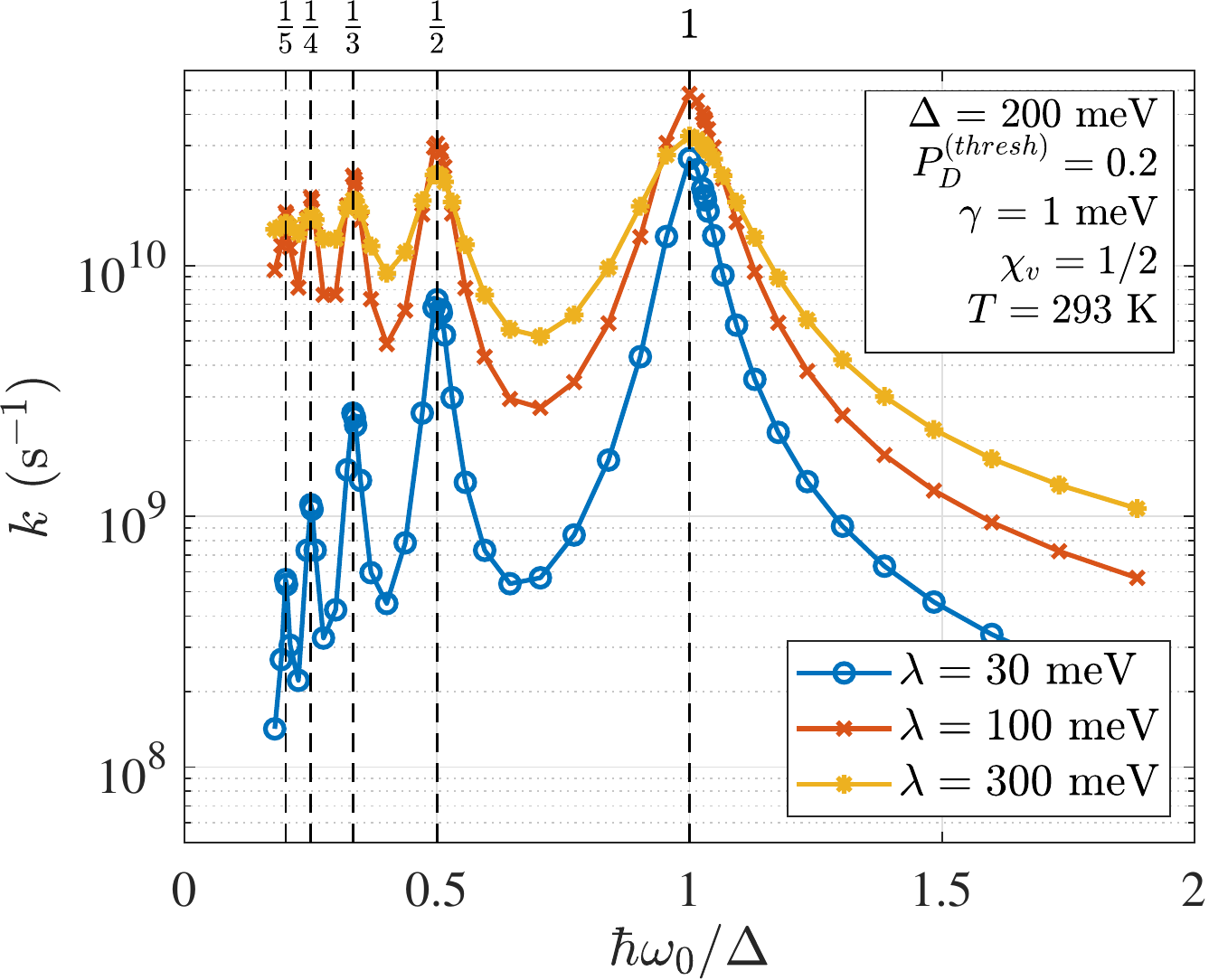}}

     \subfloat[ $\; \chi_v=2$ ]{\includegraphics[width=1\columnwidth]{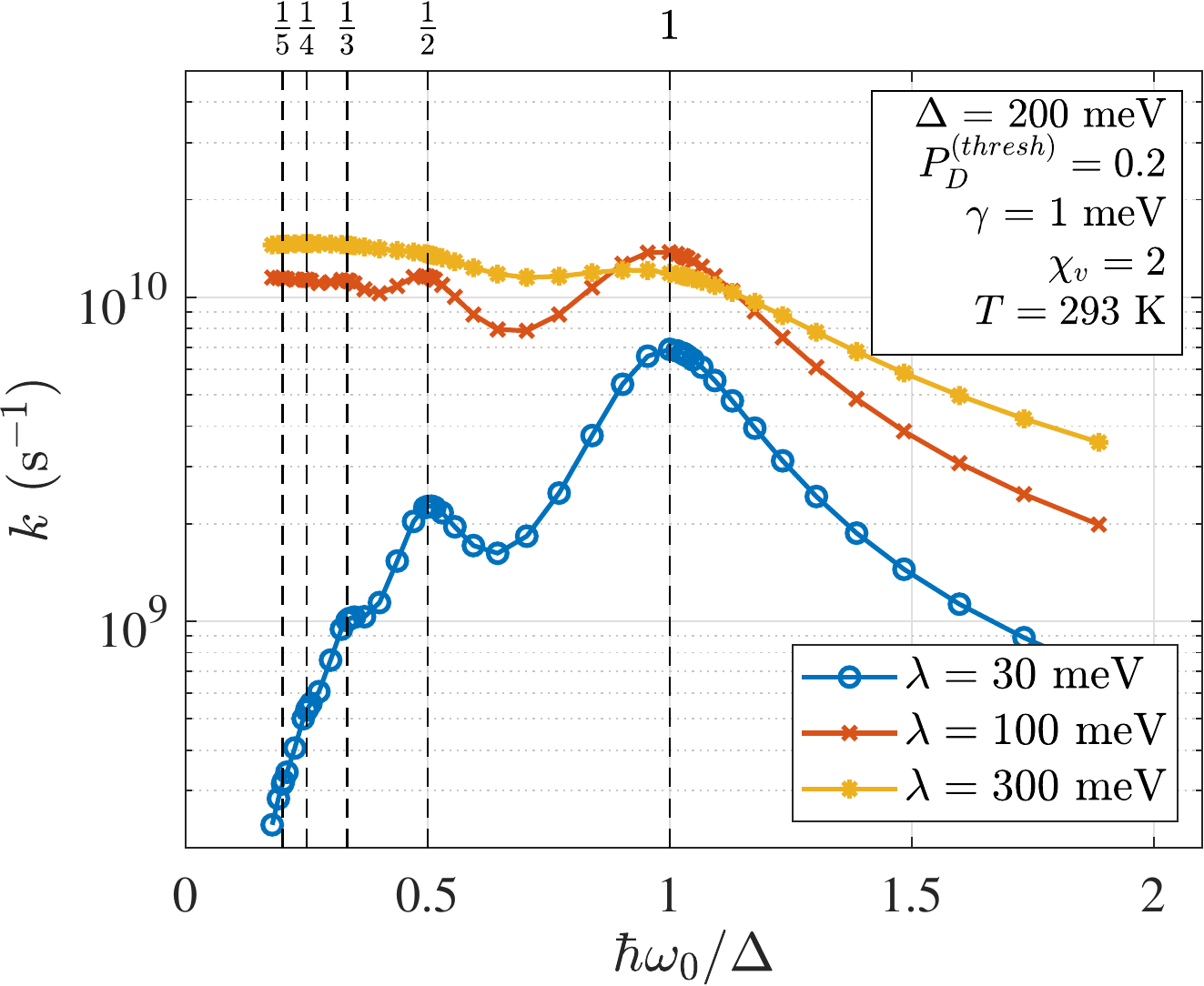}}
\caption{Resonant peaks in $k \Argum{f_o}$ are broadened as odorant reorganization energy, $\lambda$, is increased. Also, increasing $\lambda$ contributes to a higher ET rate, $k$, for lower frequencies. This is seen in subfigure (a) for a weaker odorant-environment coupling [$\chi_v = 1/2$], and again in subfigure (b) for a stronger coupling [$\chi_v = 2$].}
  \label{fig:compare_lambda}
\end{figure}
% ------------------------------------------------------------   
  
% ------------------------------------------------------------
The absence of the odorant is modeled in the limit of $\lambda \rightarrow 0$. Since coupling between the ET event and the environment already is negligible, the electronic system is unable to dissipate energy. Thus, the $\ket{D}\rightarrow \ket{A}$ transition becomes less likely because of the non-negligible detuning $\Delta$, and rate decreases to zero. This is seen in Figure \ref{fig:compare_lambda}. On the other hand, increasing $\lambda$ strengthens the electron-vibration coupling. An excessively strong coupling (high $\lambda$) broadens the peaks in $k \Argum{\hbar \omega_o/\Delta}$.  
 Figure\ \ref{fig:compare_lambda} shows that increasing the reorganization energy $\lambda$ contributes to a higher electron transfer rate $k$ for lower frequencies. Resonant peaks are broadened as $\lambda$ increases. The $\lambda$-dependence of rate $k$ seen here is appears to be consistent with Fermi's golden rule.

% ------------------------------------------------------------
\subsubsection{Tunneling energy, $\gamma$  }
\begin{figure}[hpbt]
\centering
     \subfloat[~$\chi_v=1/2$]{\includegraphics[width=1\columnwidth]{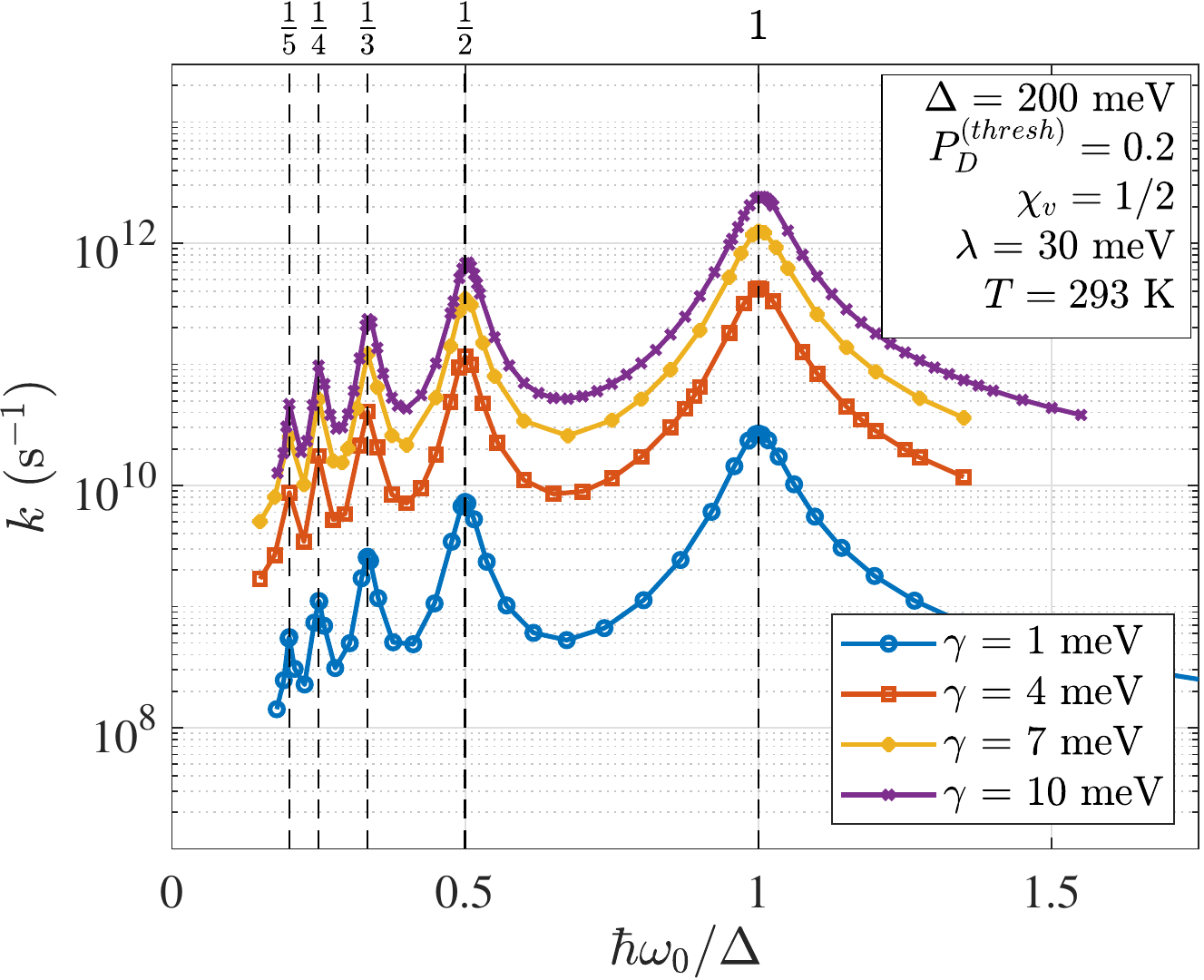}}
     
     \subfloat[ ~$\chi_v=2$ ]{\includegraphics[width=1\columnwidth]{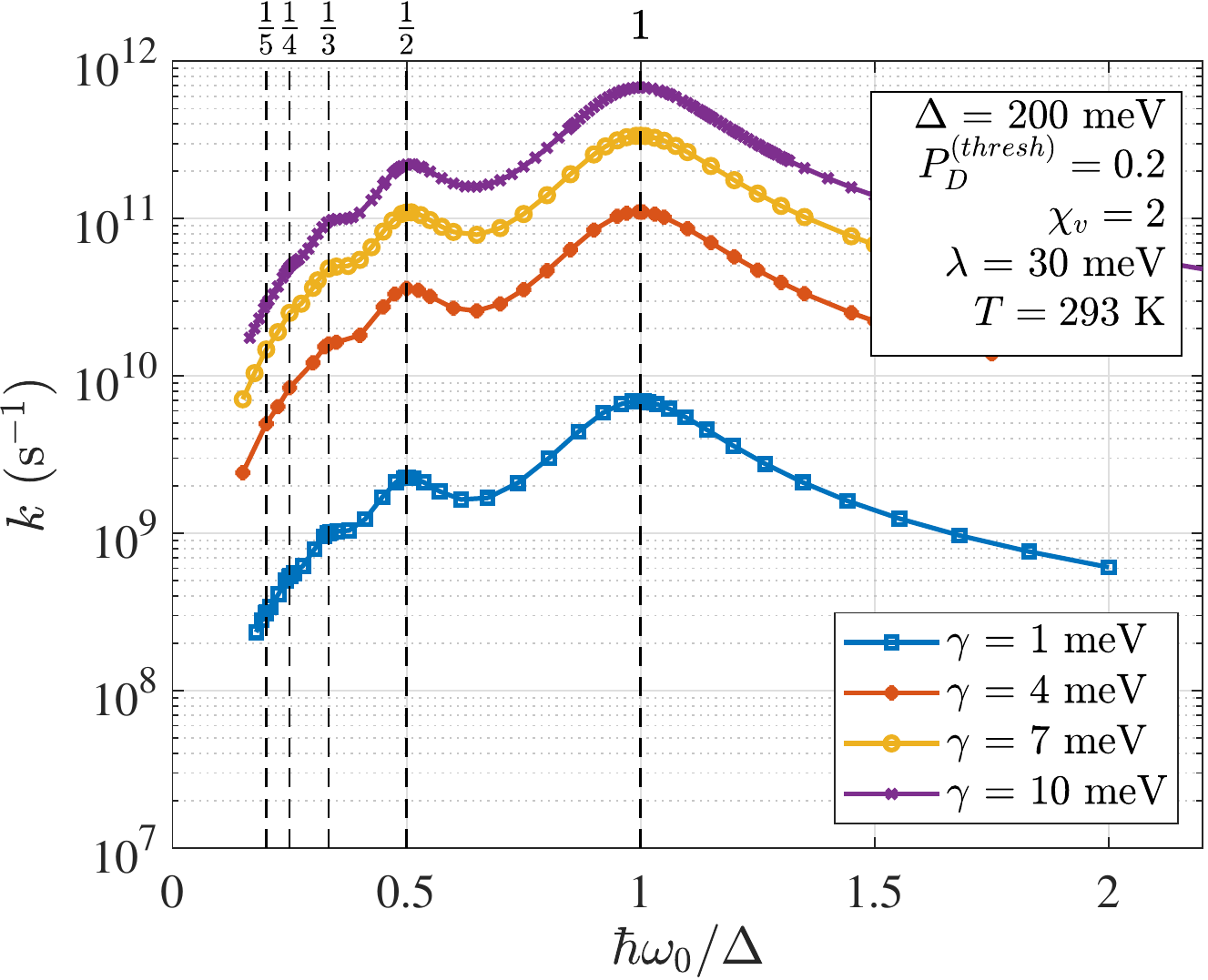}}
\caption{Increase in tunneling energy, $\gamma$, contributes to higher electron transfer rate, $k$, from state $\ket{D}$ to state $\ket{A}$. This is shown for weak odorant-environment coupling, $\chi_v=1/2$ and for strong odorant-environment coupling, $\chi_v=2$.}
  \label{fig:compare_gamma}
\end{figure}

The results of Figure\ \ref{fig:compare_gamma} 
shows that increasing the tunneling energy $\gamma$ enables a higher electron transfer rate. This is consistent with the fact that the upper speed limit of electron transfer is the Rabi oscillation frequency $\pi \gamma/\hbar$. Additionally, the $\gamma$-dependence of rate $k$ seen here appears to be consistent with Fermi's golden rule.

%  % ------------------------------------------------------------%%
%  % ------------------------------------------------------------%%%
 
\subsubsection{Temperature, $T$} \label{subsect:Temperature}
 
   % ------------------------------------------------------------  
   \begin{figure}[htbp]
      \centering
      \includegraphics[width=1\columnwidth]{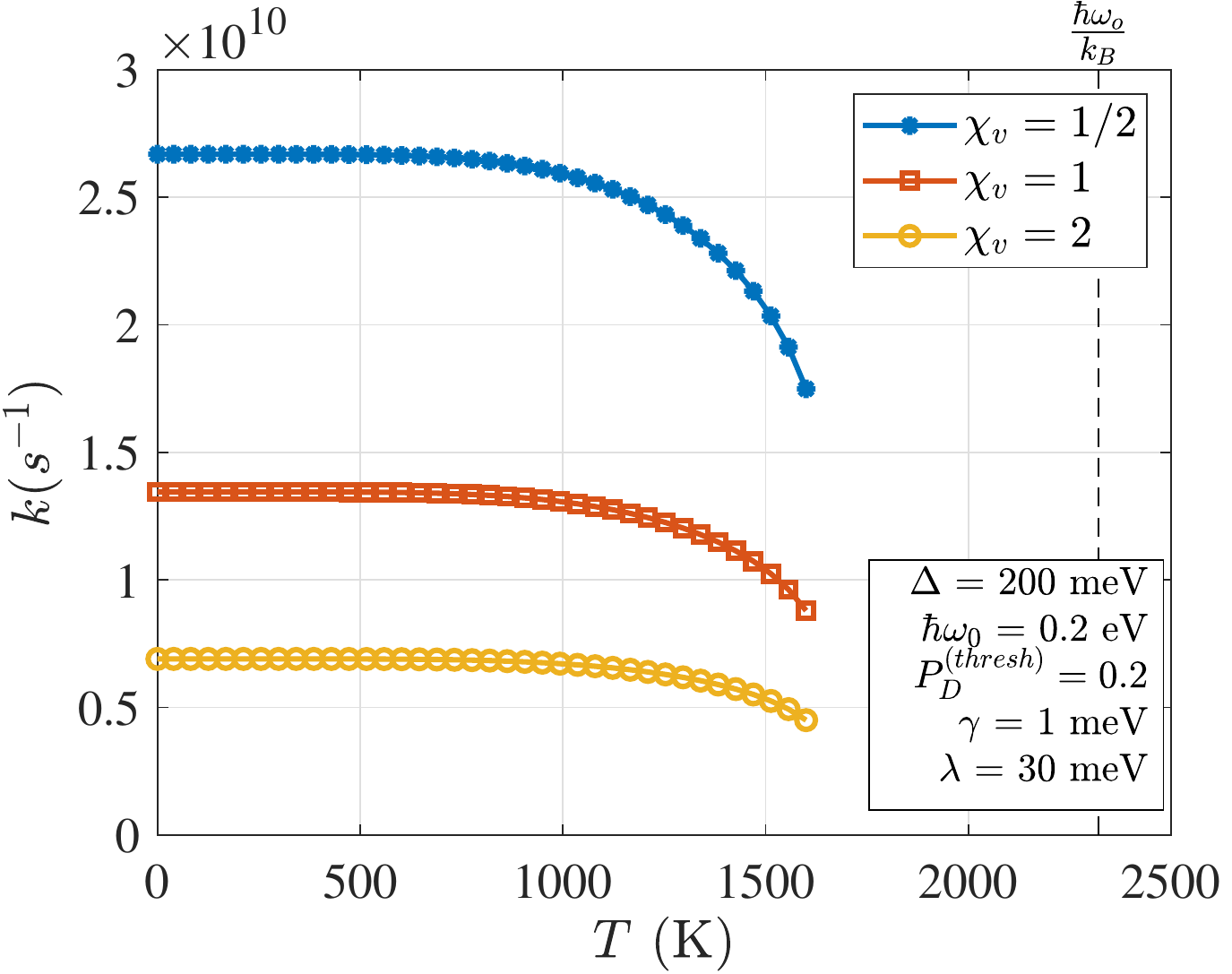}
      \caption{Electron transfer rate, $k$, from state $\ket{D}$ to state $\ket{A}$ is largely temperature-independent at terrestrial temperatures and below. ET rate is shown here as a function of environmental temperature $T$ for different values of odorant-environment coupling $\chi_v$. When $T$ is sufficiently large that $k_B T \sim \hbar \omega$, ET rate decreases since thermal environmental fluctuations enable $\ket{A}\rightarrow \ket{D}$ excitation. }
      \label{fig:k_vs_temp_odorant_parameters}
   \end{figure}
% ------------------------------------------------------------

% ------------------------------------------------------------
\begin{figure}[bhtp]
      \centering
      \includegraphics[width=1\columnwidth]{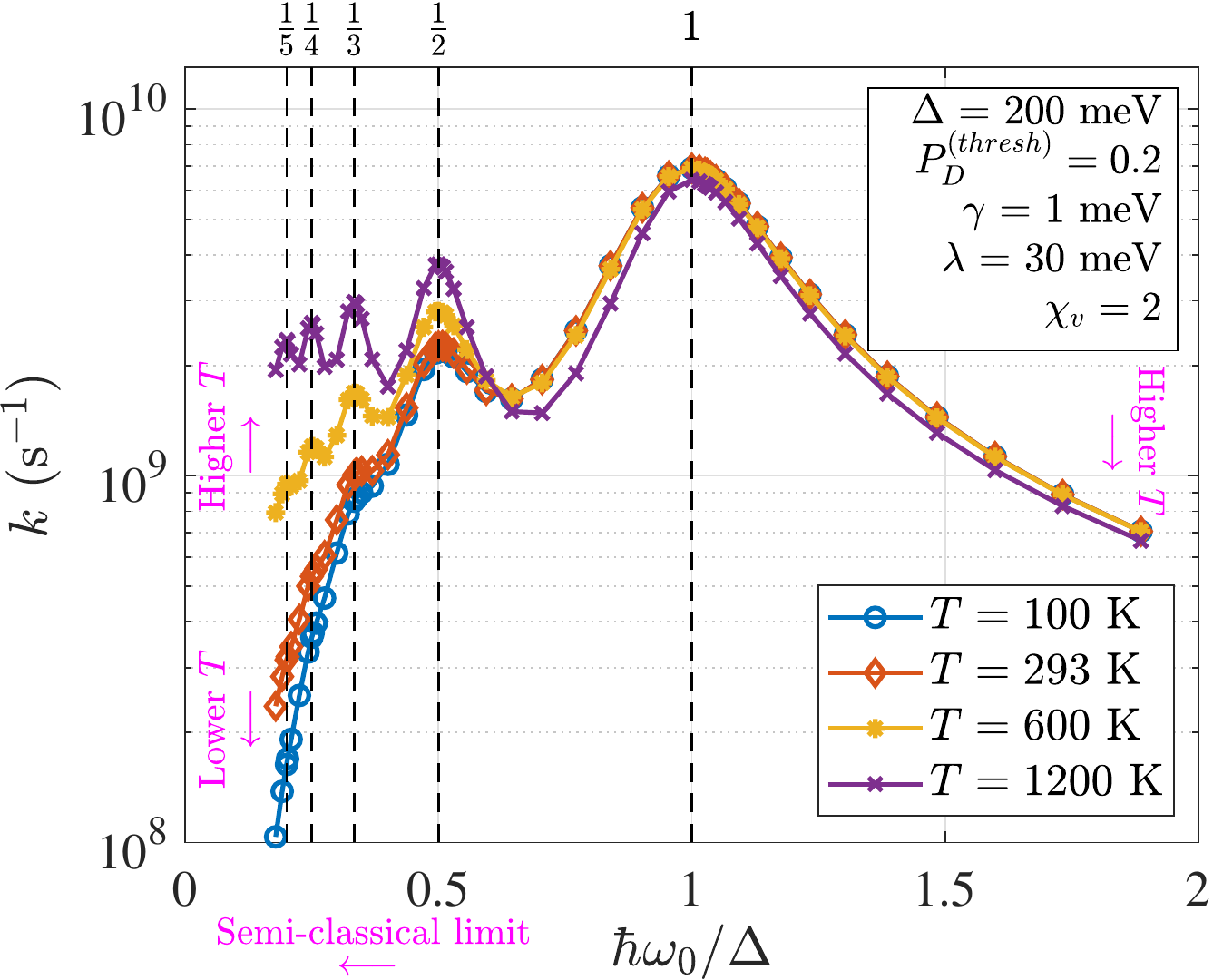}
     \caption{Model ET rates exhibit both quantum and classical/semi-classical behaviors over the spectrum of odorant frequencies $f_o$. In the low frequency limit ($\hbar \omega_o \ll \Delta$), increasing temperature increases the rate $k$ in a behavior consistent with a classical or semi-classical system in which thermal excitation allows the system to surmount a classical barrier. In the high-frequency limit ($\hbar \omega_o \gg \Delta$), ET rate, $k$, decreases with a significant increase in temperature. This is consistent with the behavior noted in Section \ref{subsect:Temperature}.}  
     \label{fig:comp_temp_strong_coupling}
   \end{figure}
  
% ------------------------------------------------------------

Figure \ref{fig:k_vs_temp_odorant_parameters} shows ET rate $k$ as a function of temperature. In the zero-temperature limit, a non-zero $k$ reveals the quantum nature of the system, as a classical or semi-classical prediction for the ET rate would be zero at in the absence of thermal excitation. For terrestrial and biological temperatures $T < 300~\mbox{K}$, ET rate $k$ is largely constant. As $T$ increases to high temperatures ($k_B T \sim \hbar \omega_o $), $k$ begins to decrease. This is because as $T$ increases, the system is increasingly thermally excited, raising the probability $P_D \Argum{t}$ of finding the electron on the $D$ site.  
This increases the time $t_{ET}$ it takes for $P_D \Argum{t}$ to decay to the threshold defined in our calculation, thus decreasing $k$. While the temperature-dependence of $k$ from this model is largely consistent with Fermi's golden rule, a point of divergence between the two models is rate behavior at low temperature. Fermi's golden rule predicts $k=0$ at $T=0$; however, the present model predicts $k \neq 0$ at $T=0$.

Interestingly, a quantum-to-classical transition emerges when we consider how temperature $T$ and oscillator frequency $\omega_o$ affect rate $k$. Figure\ \ref{fig:comp_temp_strong_coupling} shows calculations of $k$ for various values of $\omega_o$ and $T$. For low $\omega_o$, reducing $T$ decreases the ET rate $k$. This result is consistent with classical and semi-classical rate models, which require thermal excitation to cross a potential barrier. For low $\omega_o$, the fully-quantum system gives rise to this semi-classical behavior because lowering $\omega_o$ reduces the spacing between eigenvalues of the fully-quantum system (see Figure \ref{fig:marcus12}). For low-enough $\omega_o$, the energy quantization in the fully-quantum model is small compared to the barrier height between states $\ket{D}$  and $\ket{A}$ in the semi-classical system. Thus, the semi-classical behaviors of the system will be dominant. On the other hand, when $\omega_o$ is large, energy quantization in the system is significant, the system manifests more quantum mechanical behaviors. Here, increasing $T$ only slightly decreases $k$, as discussed previously. With large $\omega_o$, a high rate $k$ persists, even at low temperature, $T = 0$. This is a truly non-classical result.

\subsubsection{Discrimination between isotopomers}
%%% ---------------------------------------------------------%%%%%
      \begin{figure}[phbt]
\centering
     \subfloat[~$\chi_v=1/2$]{\includegraphics[width=1\columnwidth]{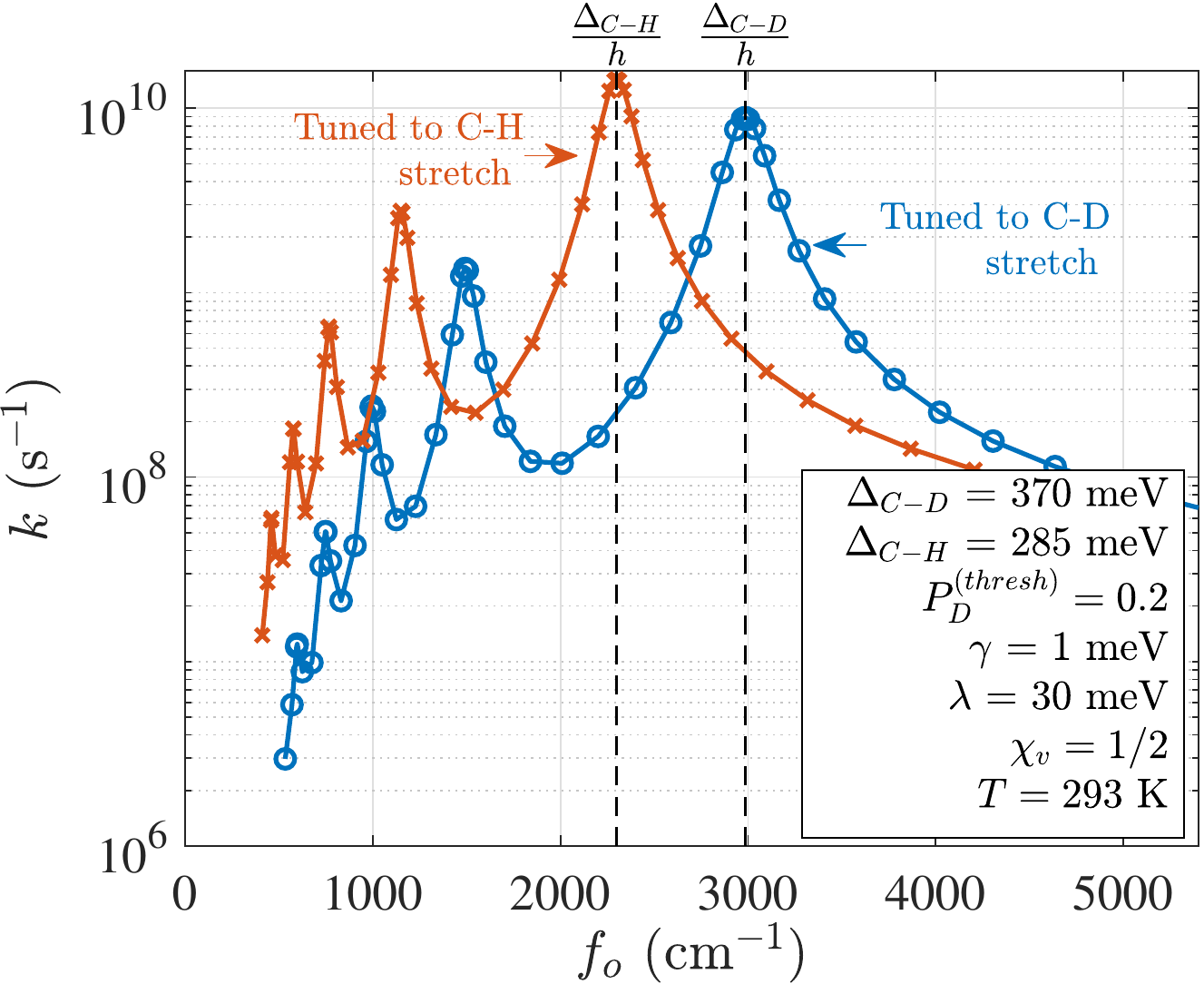}}
     
     \subfloat[~$\chi_v=2$]{\includegraphics[width=1\columnwidth]{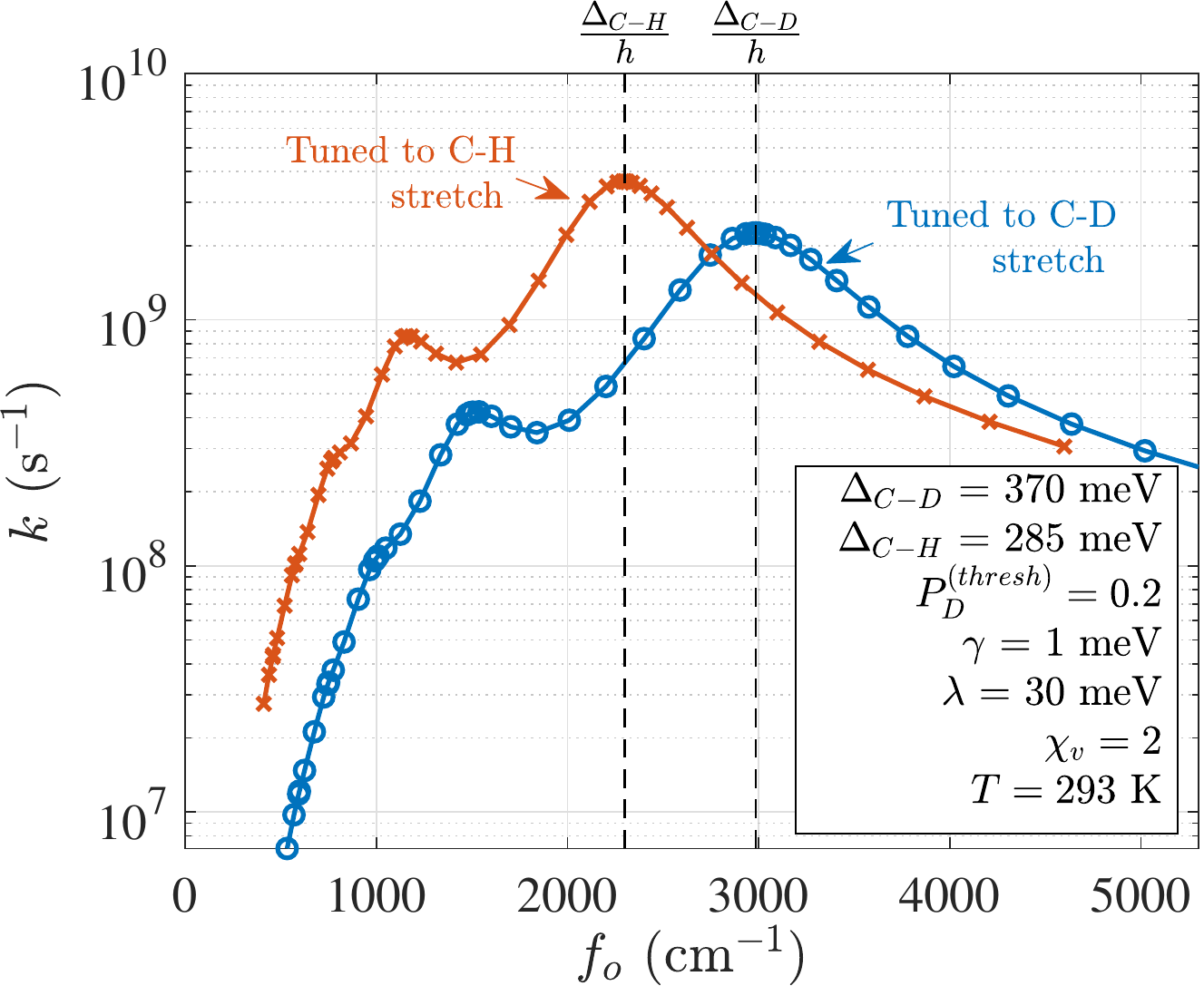}}
\caption{The model developed here exhibits a differential response to isotopomers. (a) A receptor tuned to the C-H stretch responds to the C-H stretch with a rate $k$ that is more than one order of magnitude higher than the rate due to the presence of a C-D stretch vibrational mode. A receptor tuned to the C-D stretch exhibits a similarly higher ET rate when an odorant with a C-D ligand is present than when an odorant with a C-H ligand is present. (b) Peak broadening occurs with stronger odorant-environmental coupling, and the change in rate between isotopomers is reduced for each receptor.}
  \label{fig:cdch}
\end{figure}
 %%% ---------------------------------------------------------%% 

Next, the model developed here is applied to the study of receptor selectivity in the odorant-receptor complex response to isotopomers. Specifically, we considered a receptor's ability to spectroscopically distinguish acetophenone from its fully-deuterated isotopomer acetophenone-d8. Here, we assume that the C-H and C-D stretch play the distinguishing role.\cite{2011FruitFly,2012QOlfactionReview}

Figure \ref{fig:cdch} shows the spectroscopic response of $k$ for two receptors: one receptor is resonantly tuned to the C-H stretch frequency $f_o ^{(C-H)} = 2300~\mbox{cm}^{-1}$ by setting $\Delta = \Delta_{C-H} = \hbar \omega_o ^{(C-H)}$, and another is resonantly tuned to the C-D stretch frequency $f_o^{(C-D)} = 3000~\mbox{cm}^{-1}$ by setting $\Delta = \Delta_{C-D} = \hbar \omega_o^{(C-D)}$. The response of each receptor is calculated for various values of $\omega_o$, and the strongest resonant peak occurs as expected for the preferred odorant when $\Delta = \hbar \omega_o$. For each receptor, if odorant-environment coupling is weak ($\chi_v = 1/2$), ET rate $k$ drops by almost two orders of magnitude when the non-preferred isotopomer is in the receptor.  For a stronger odorant-environment coupling ($\chi_v = 2$), spectral broadening results in the spectroscopic $k \Argum{\omega_o}$ response, and rate drops by more than 50\% when the preferred molecule is replaced by its isotopomer. Increased environmental coupling reduces the receptor's selectivity to its preferred odorant.

  % ------------------------------------------------------------%
  
\subsubsection{Power dissipation}

% ------------------------------------------------------------%
 \begin{figure}[bthp]
      \centering
      \includegraphics[width=1\columnwidth]{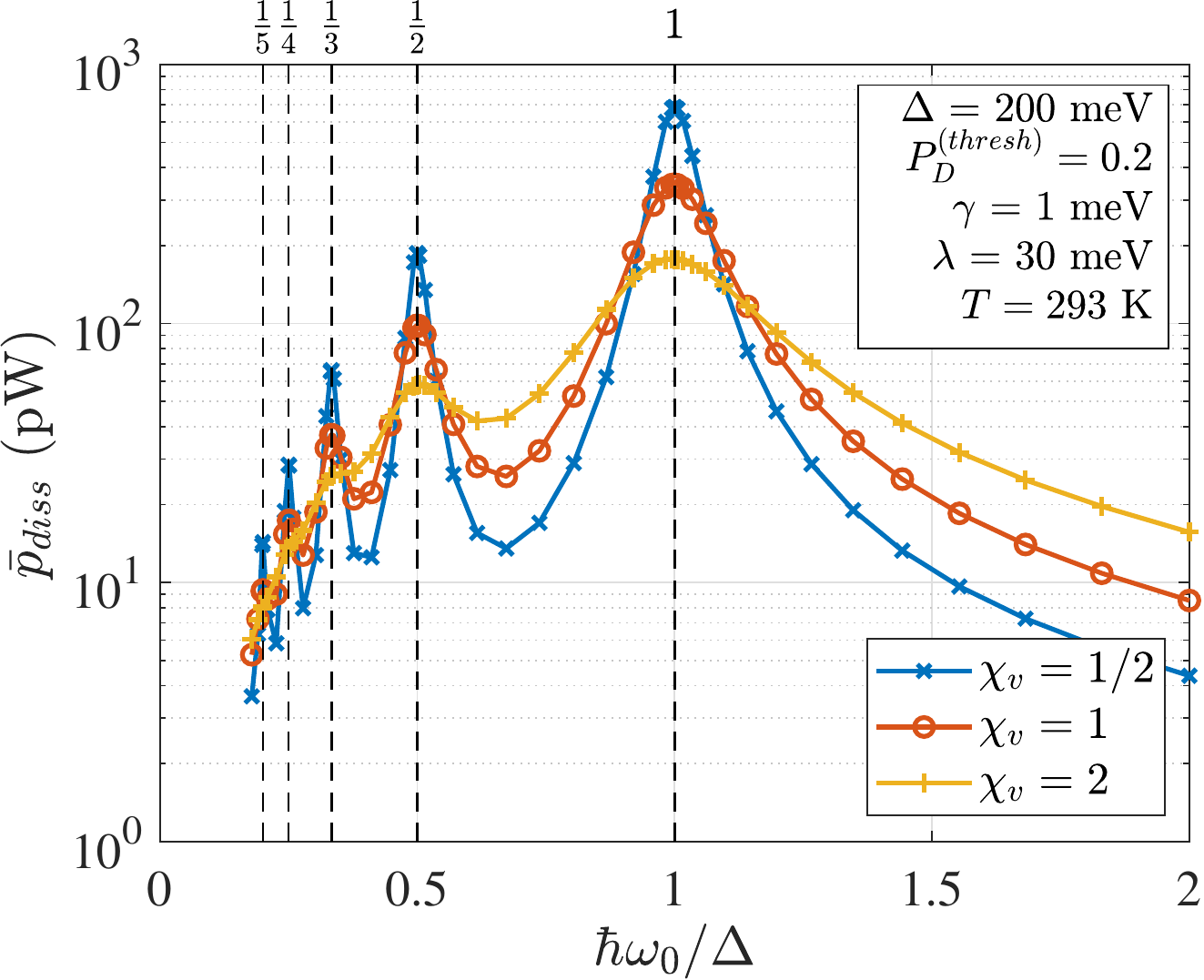}
    \caption{Power dissipation $\bar{p} \TextSub{diss}$ to the environment exhibits a frequency-dependence very similar to the the frequency dependence of rate $k$ to odorant frequency (compare with Figure \ref{fig:compare_coupling}). Peaks for $\bar{p} \TextSub{diss}$ occur when $\Delta = s \hbar \omega_o$ for some integer $s$, and resonant peaks broaden with stronger odorant-environment interaction (increasing $\chi_v$).} 
      \label{fig:power_compare_coupling2}
   \end{figure} 
   % ------------------------------------------------------------%

The dissipation of power and energy is essential to a $\ket{D} \rightarrow \ket{A}$ transition. A plot of average power dissipation $\bar{p} \TextSub{diss}$ (see Figure\ \ref{fig:power_compare_coupling2}) for various odorant frequencies shows that the system dissipates the highest rates of average power at frequencies satisfying $\Delta = s \hbar \omega_o $ for some positive integer $s$. This coincides exactly with the resonant peaks in rate (see Figure\ \ref{fig:compare_coupling}), underscoring the enabling role dissipation plays in this process: the more quickly the electron can dissipate power, the faster it can make the $\ket{D} \rightarrow \ket{A}$ transition, increasing the ET rate. This relationship is a direct consequence of the conservation of energy, and the ET is thus facilitated by its ability to relax by $\Delta$ via the excitation of odorant phonons or by dissipating power to the thermal environment.

\begin{figure}[bthp]
      \centering
      \includegraphics[width=1\columnwidth]{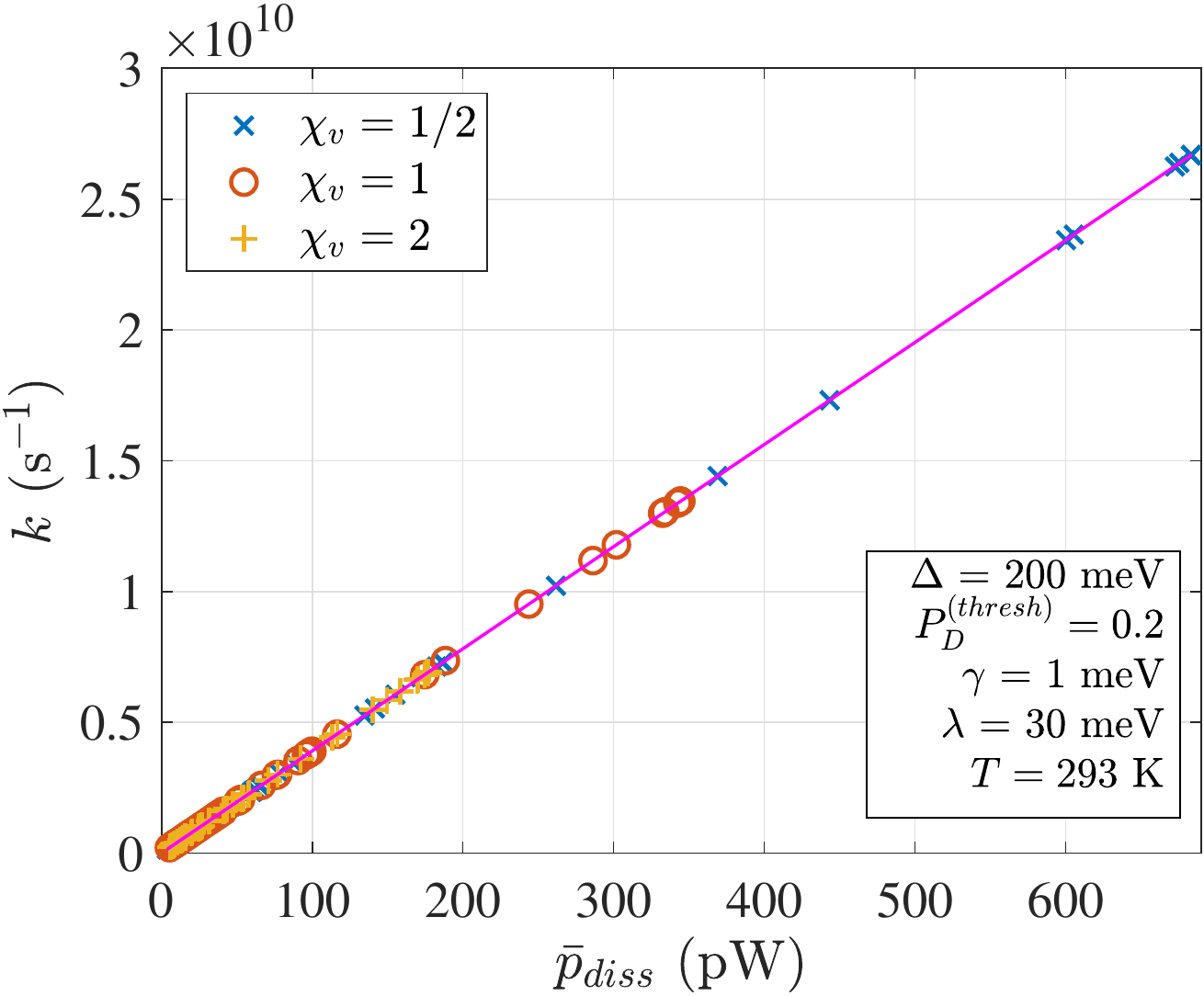}
    \caption{Power dissipation $\bar{p} \TextSub{diss}$ exhibits a linear correlation with ET rate. This scatter plot of rate data from Figure \ref{fig:compare_coupling} and power dissipation data from Figure \ref{fig:power_compare_coupling2} shows a highly linear relationship between the two sets of data. A fitting line is drawn in purple, and the data lies along this same line for several odorant-environment coupling strengths.} 
      \label{fig:k_vs_power_cartesian}
   \end{figure}

The relationship between rate $k$ and $\bar{p} \TextSub{diss}$ exhibits a highly linear relationship, as seen in the scatter plot of Figure \ref{fig:k_vs_power_cartesian}. This scatter plot shows the correlation between the rate data from Figure \ref{fig:compare_coupling} and the power dissipation data of Figure \ref{fig:power_compare_coupling2}. One regression line is remarkably effective for three different odorant-environment coupling strengths ($\chi_v \in \{ 0.5, 1, 2\}$). 

% ------------------------------------------------------------ 

\begin{figure}[phbt]
\centering
     \subfloat[ ~ $\chi_v=1/2$ ]{\includegraphics[width=1\columnwidth]{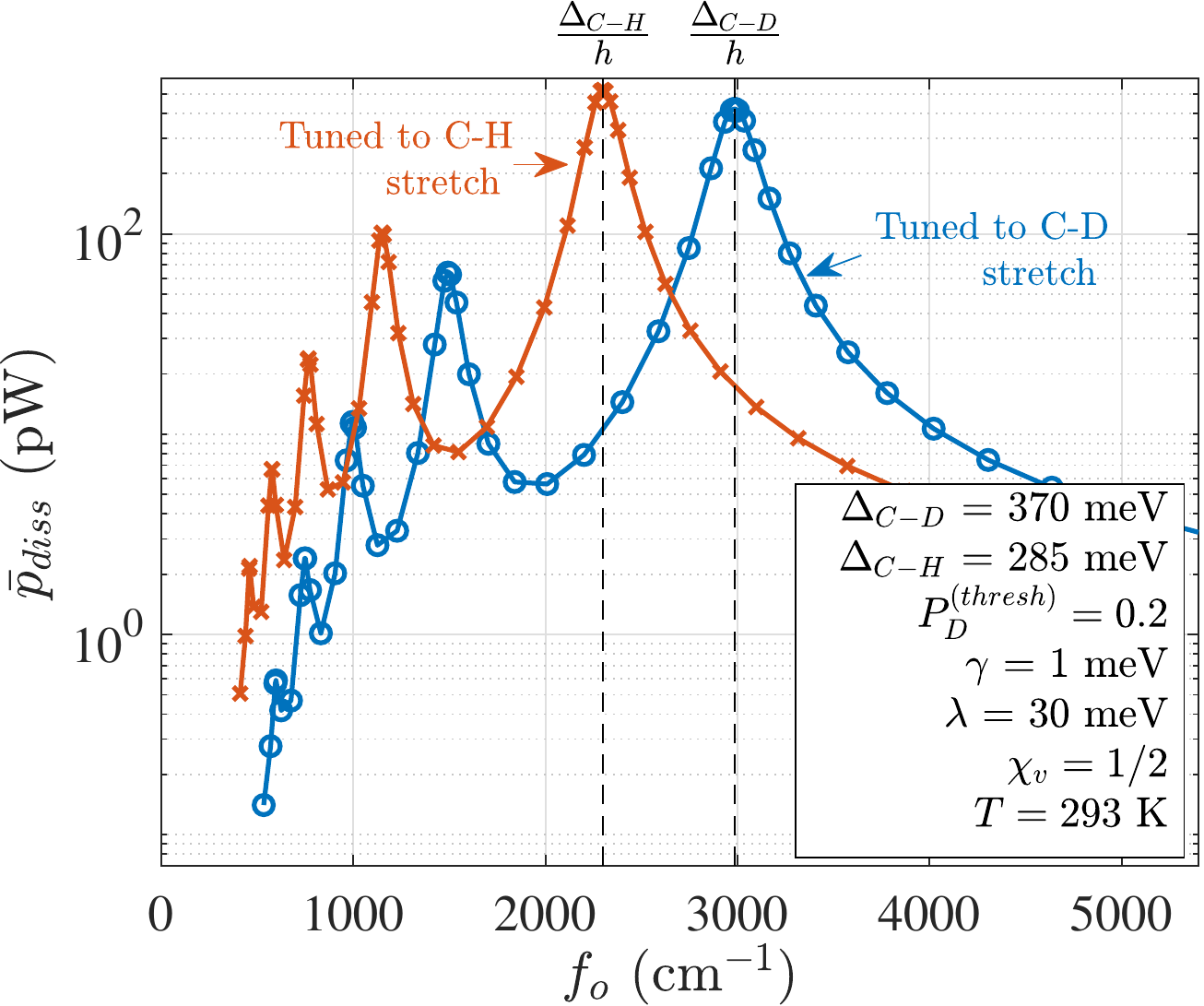}}
     
     \subfloat[ ~ $\chi_v=2$ ]{\includegraphics[width=1\columnwidth]{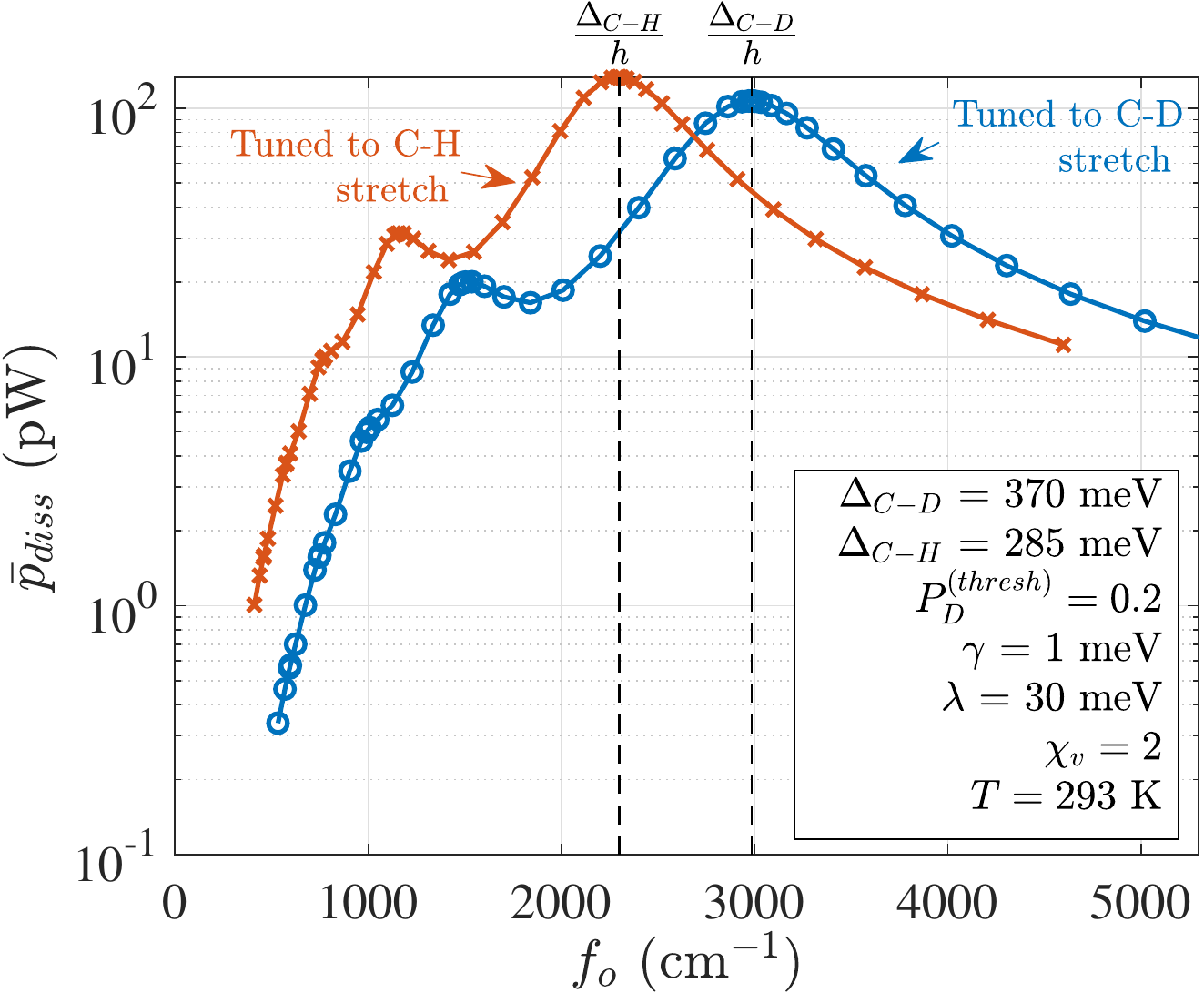}}
\caption{Average power dissipation to the environment is frequency-selective\textemdash and therefore ligand-selective. Power dissipation peaks when a receptor tuned to the C-H (or C-D) stretch couples to a ligand with a frequency $f_o$ matching the frequency of the C-H (or C-D) stretch.}
  \label{fig:power_cdch}
\end{figure}

% ------------------------------------------------------------    
Finally, we use power dissipation to quantify the selectivity of a receptor within this model to its preferred isotopomer. Power dissipation $\bar{p} \TextSub{diss}$ is plotted as a function of odorant frequency $f_o$ in Figure \ref{fig:power_cdch}. The power dissipation curves of Figure \ref{fig:power_cdch} are very similar to the rate-versus-frequency curves for the same receptors (see Figure \ref{fig:cdch}).

We quantify receptor selectivity by taking the ratio of power dissipation at the characteristic vibrational frequency of the preferred odorant to the rate of power dissipation at the frequency characteristic of the non-preferred odorant. For the receptor tuned to the C-H (or C-D) stretch mode, we use power dissipation in the presence of a C-H (or C-D) bond and divide by the power dissipation in the presence of a C-D (or C-H) stretching mode. With an odorant-environment coupling of $\chi_v = 1/2$, selectivity for the preferred odorant is $\sim15~\mbox{dB}$ (see Table \ref{table:selectivity_weak}). Stronger odorant-environment coupling ($\chi_v=2$) broadens the peaks in the $\bar{p} \TextSub{diss} \Argum{f_o}$ curve, and selectivity drops to $\sim 5~\mbox{dB}$ (see Table \ref{table:selectivity_strong}).
 
% ------------------------------------------------------------ 

% ------------------------------------------------------------ 
\begin{table}[htbp]
\centering
\begin{tabular}{||p{0.7in}|p{0.5in}|p{0.5in}|p{.7in}||}
\hline
\multirow{3}{0.7in}{\textbf{Receptor tuned to}} & \multicolumn{2}{p{1in}|}{\textbf{Power dissipation (pW)}} & \multirow{3}{0.7in}{\textbf{Selectivity (dB)}}  \\
\cline{2-3} 
 & \textbf{C-H Stretch} & \textbf{C-D Stretch} &  \\
\hline\hline
 C-H stretch& 521.7 & 17.2 & 14.81 \\
\hline
 C-D stretch& 10.6 & 420.7 & 16 \\
\hline
\end{tabular}
\caption{Power dissipation from the receptor-odorant complex to the environment is enhanced when a ligand is present for which the receptor is tuned (i.e., $\Delta = \hbar \omega_o$). Here, an odorant-environment coupling  of $\chi_v = 1/2$ is assumed. When the preferred ligand is replaced by its related isotopomer, $\Delta \neq \hbar \omega_o$, and power dissipation is suppressed by $\sim 15.4~\mbox{dB}$. This frequency-selective dissipation is used to quantify the selectivity of the receptor to a particular vibrational mode over in a related isotopomer.}
\label{table:selectivity_weak}
\end{table}
 
% ------------------------------------------------------------ 

% ------------------------------------------------------------ 
\begin{table}[htbp]
\centering
\begin{tabular}{||p{0.7in}|p{0.5in}|p{0.5in}|p{.7in}||}
\hline
\multirow{3}{0.7in}{\textbf{Receptor tuned to}} & \multicolumn{2}{p{1in}|}{\textbf{Power dissipation (pW)}} & \multirow{3}{0.7in}{\textbf{Selectivity (dB)}}  \\
\cline{2-3} 
 & \textbf{C-H Stretch} & \textbf{C-D Stretch} &  \\
\hline\hline
 C-H stretch& 133.4 & 46.1& 4.62 \\
\hline
 C-D stretch& 31.4 & 106.9& 5.32 \\
\hline
\end{tabular}
\caption{In analogy to Table \ref{table:selectivity_weak}, receptor selectivity is calculated and tabulated for stronger odorant-environment coupling ($\chi_v = 2$). Increased odorant-environment coupling reduces receptor selectivity (compare to Table \ref{table:selectivity_weak}). }
\label{table:selectivity_strong}
\end{table}
 
% ------------------------------------------------------------ -----
 \subsection{Direct (Electron-environment) Coupling Dominates}
 % ------------------------------------------------------------ 
 
\begin{figure}[bthp]
      \centering
      \includegraphics[width=1\columnwidth]{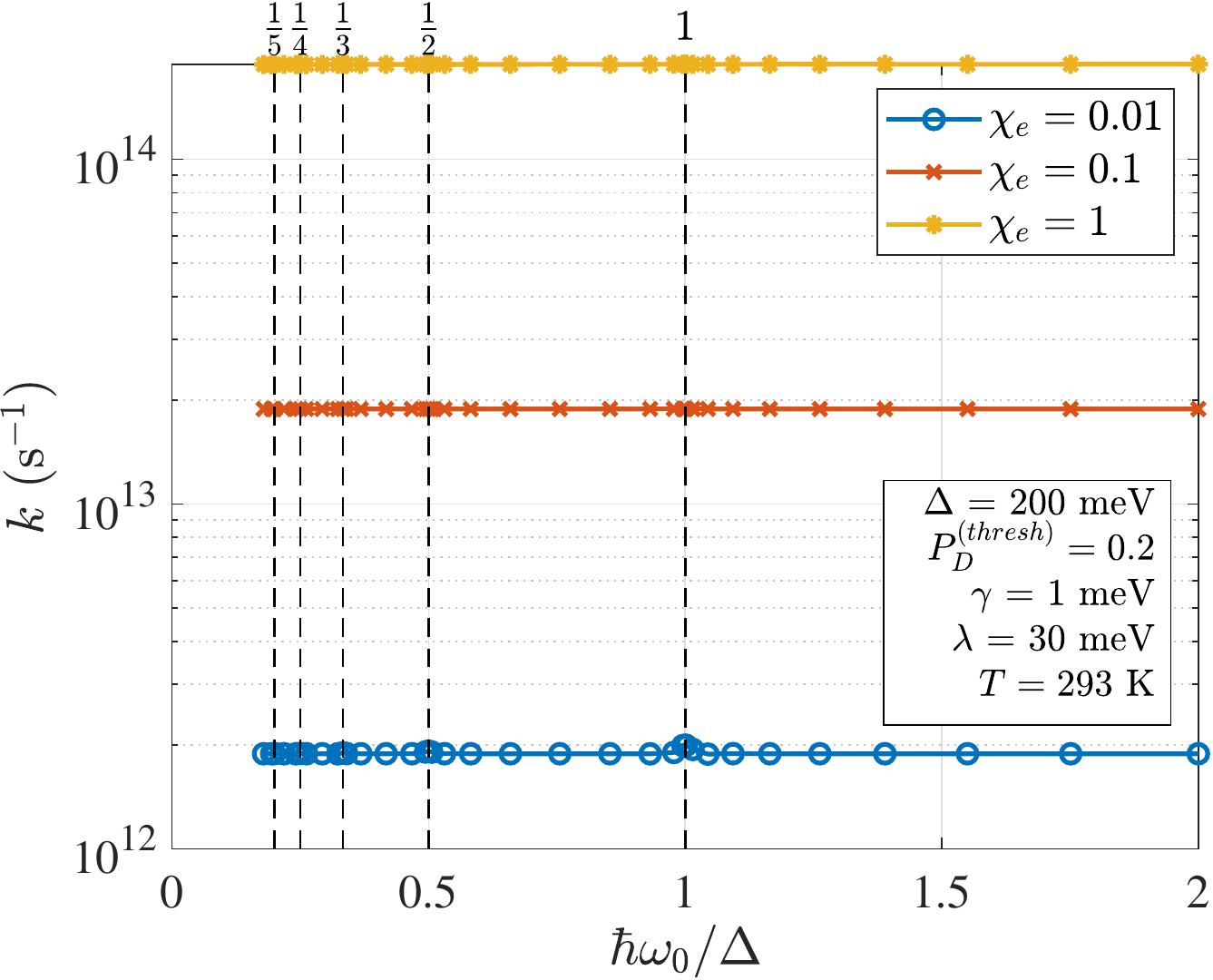}
    \caption{Resonant behaviors are almost completely suppressed when the indirect dissipation path is disrupted by severing the odorant-environment connection ($\chi_v \rightarrow 0$). Here, the rate is largely independent of odorant vibrational frequency $f_0$ and is determined solely by the strength of direct dissipation $\chi_e$. ET rate increases with increasing direct electron-environment coupling $\chi_e$.} 
      \label{fig:compare_coupling_LargeT1}
   \end{figure}

 When direct electron-environment coupling provides the strongly-dominant dissipation path ($\xi \rightarrow 0$), spectroscopic behaviors vanish, and rate becomes largely independent of the vibrational frequency of the odorant. This is seen in Figure \ref{fig:compare_coupling_LargeT1}.  Here, a raised electron-environment coupling (increasing $\chi_e$) enables a faster relaxation and a higher ET rate $k$.
 
 A very subtle resonant effect in rate is seen if $\chi_e$ is low. Slight deviations in the ET rate at resonant frequencies ($\omega_o = s \Delta/\hbar$ for some integer $s > 0$) are revealed (see data for $\chi_e = 0.01$). These deviations are due to a transient response in which a resonant condition arises, and the conservation of energy facilitates rapid power transfer from electron to an integer number of odorant phonons. This allows a faster ET: the probability $P_D \Argum{t}$ crosses the threshold $P^{(thresh)}_D$ sooner, resulting in a slightly higher rate $k$. However, since this energy cannot be dissipated from the odorant to the environment, the energy is subsequently transferred back to the electron, with some $\ket{A}\rightarrow \ket{D}$ back-transfer, and a $P_D \Argum{t}$ that had crossed below $P^{(thresh)}_D$ may cross above $P^{(thresh)}_D$ again.
 
   % ------------------------------------------------------------ 

% ------------------------------------------------------------  
 
\begin{figure}[bthp]
      \centering
      \includegraphics[width=1\columnwidth]{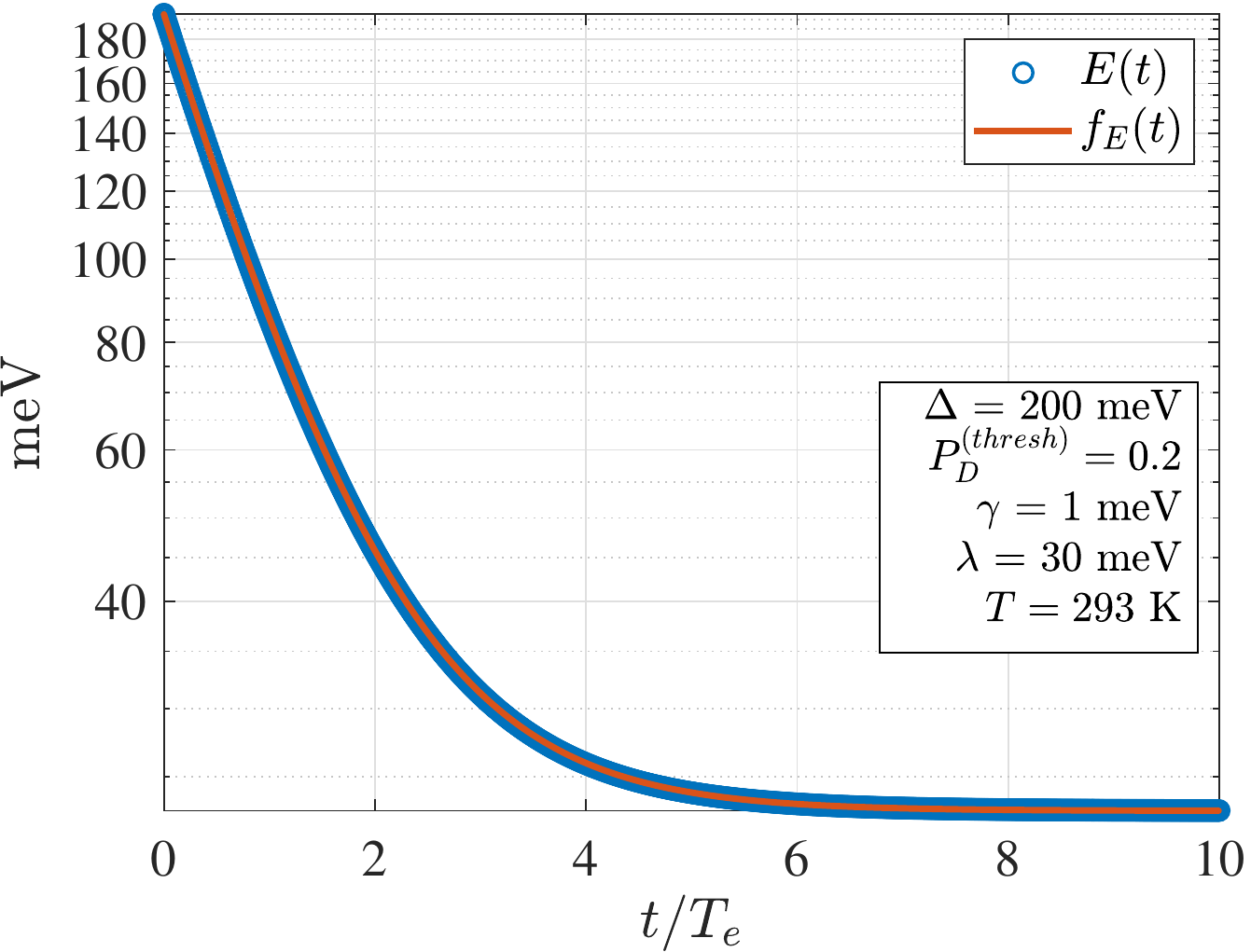}
    \caption{The direct electron-environment dissipation path drives an exponential relaxation with time constant $T_e$. Here, $T_e = 33~\mbox{fs}$, and electron+odorant total energy $E \Argum{t}$ matches an exponential fit $f_E \Argum{t}$.
    }
      \label{fig:energy_vs_t}
   \end{figure}

Figure \ref{fig:energy_vs_t} shows that the direct electron-to-environment interaction indeed drives an exponential relaxation characterized by time constant $T_e$. Here, $E\Argum{t} = \mbox{Tr} \; \MyPar{\hat{H} \hat{\rho} \Argum{t}}$, the expectation value of energy, is plotted with an exponential fit, given by
 \begin{equation}
f_E \Argum{t} = E \Argum{\infty} + \left(  E \Argum{0} -  E \Argum{\infty} \right) \left(1 - e^{-t/T_e} \right) \; .
\end{equation}
The calculated $E\Argum{t}$ from the model matches the exponential $f_{E} \Argum{t}$ very precisely.
 
 % ------------------------------------------------------------ 

% ------------------------------------------------------------  

 % ------------------------------------------------------------ -----
 \subsection{Both Direct and Indirect Coupling are Significant}
\begin{figure}[phbt]
\centering
     \includegraphics[width=1\columnwidth]{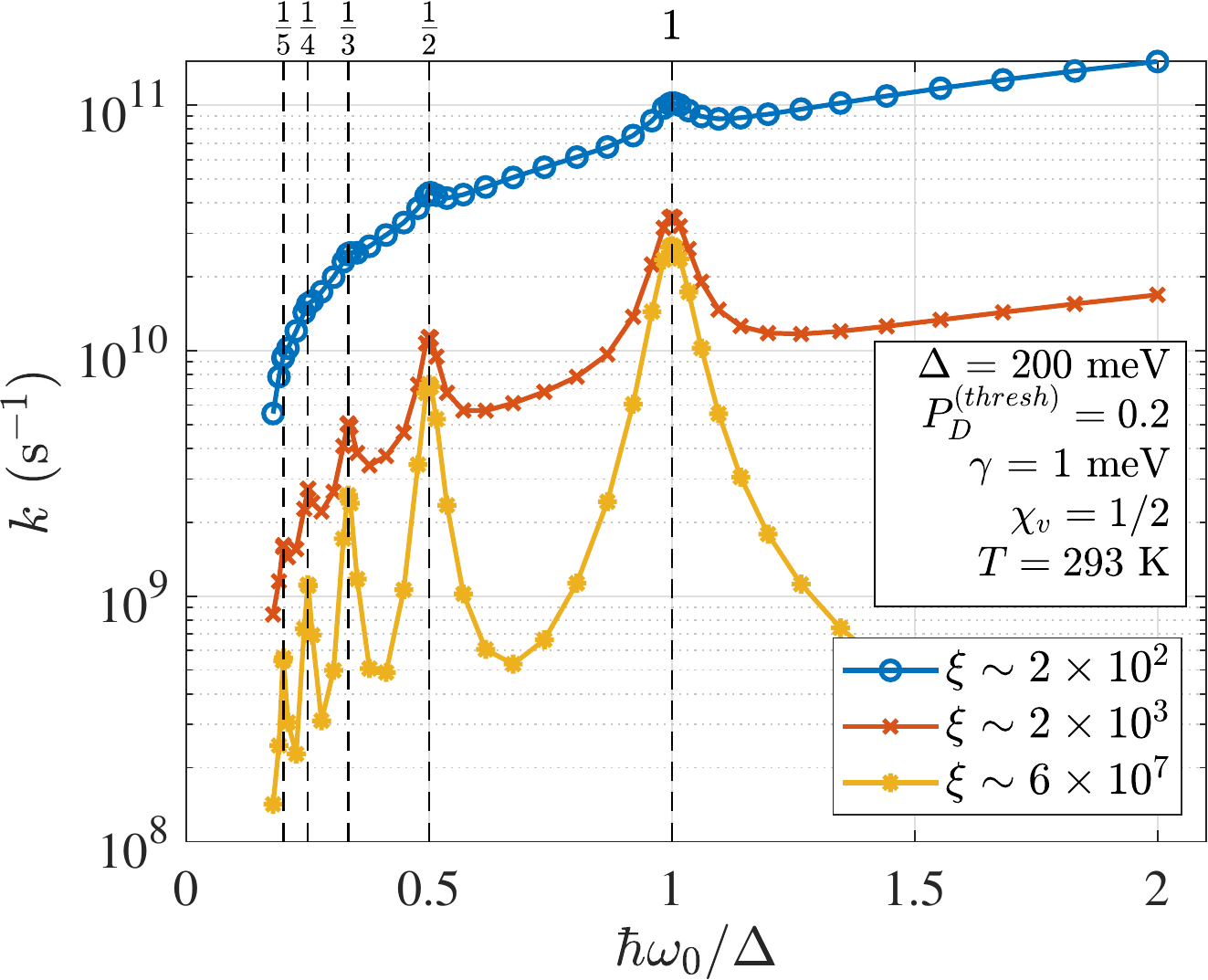}
\caption{As the strength of direct coupling is increased ($\xi$ decreases), the resonant peaks broaden, and spectroscopic behaviors in the ET rate become masked.}
  \label{fig:compareTe}
\end{figure}

 % ------------------------------------------------------------
 Here, we explore the case in which neither the direct nor indirect dissipative pathways are negligible. In Figure \ref{fig:compareTe}, the resonant peaks in the rate $k$ become less pronounced as the system transitions from an indirect-coupling-dominated (larger $\xi$) to direct-coupling-dominated (smaller $\xi$). This is because the additional (direct) dissipative pathway assists the electron transfer, raising the ET rate most noticeably in the non-resonant regions ($\Delta \neq s \hbar \omega_0$ for positive integers $s$) where ET was most highly suppressed with high $\xi$ (compare Figure \ref{fig:compare_coupling} to the red and blue traces of Figure \ref{fig:compareTe}). If $\xi$ becomes small enough, the spectroscopic response of $k(f_0)$ vanishes as the direct dissipative pathway dominates and the resonant behavior of the indirect dissipation path is masked.

% -----------------------------------------------------------------
\begin{figure}[phbt]
\centering
     \subfloat[Receptor tuned to C-D ]{\includegraphics[width=1\columnwidth]{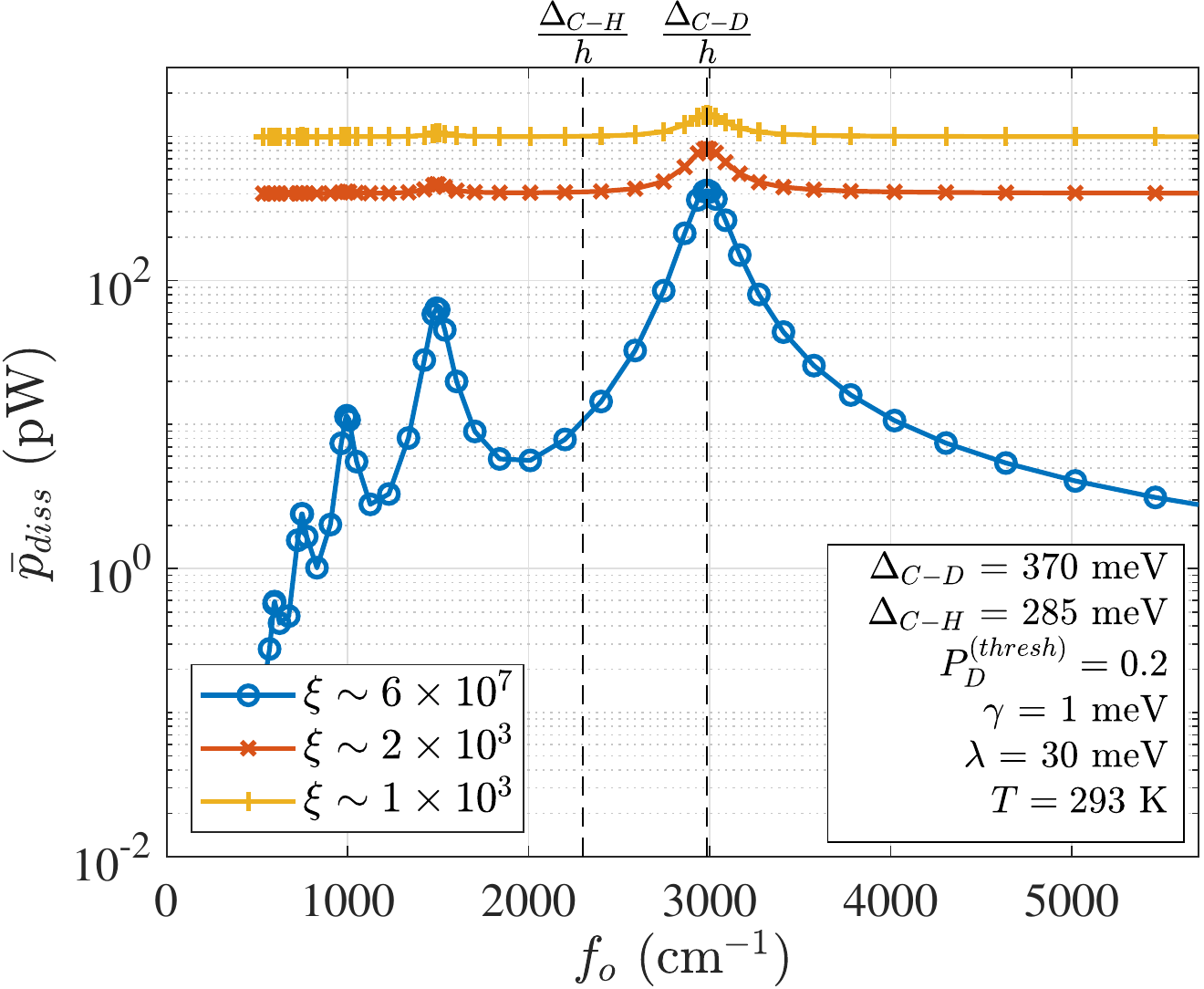}}
     
     \subfloat[Receptor tuned to C-H ]{\includegraphics[width=1\columnwidth]{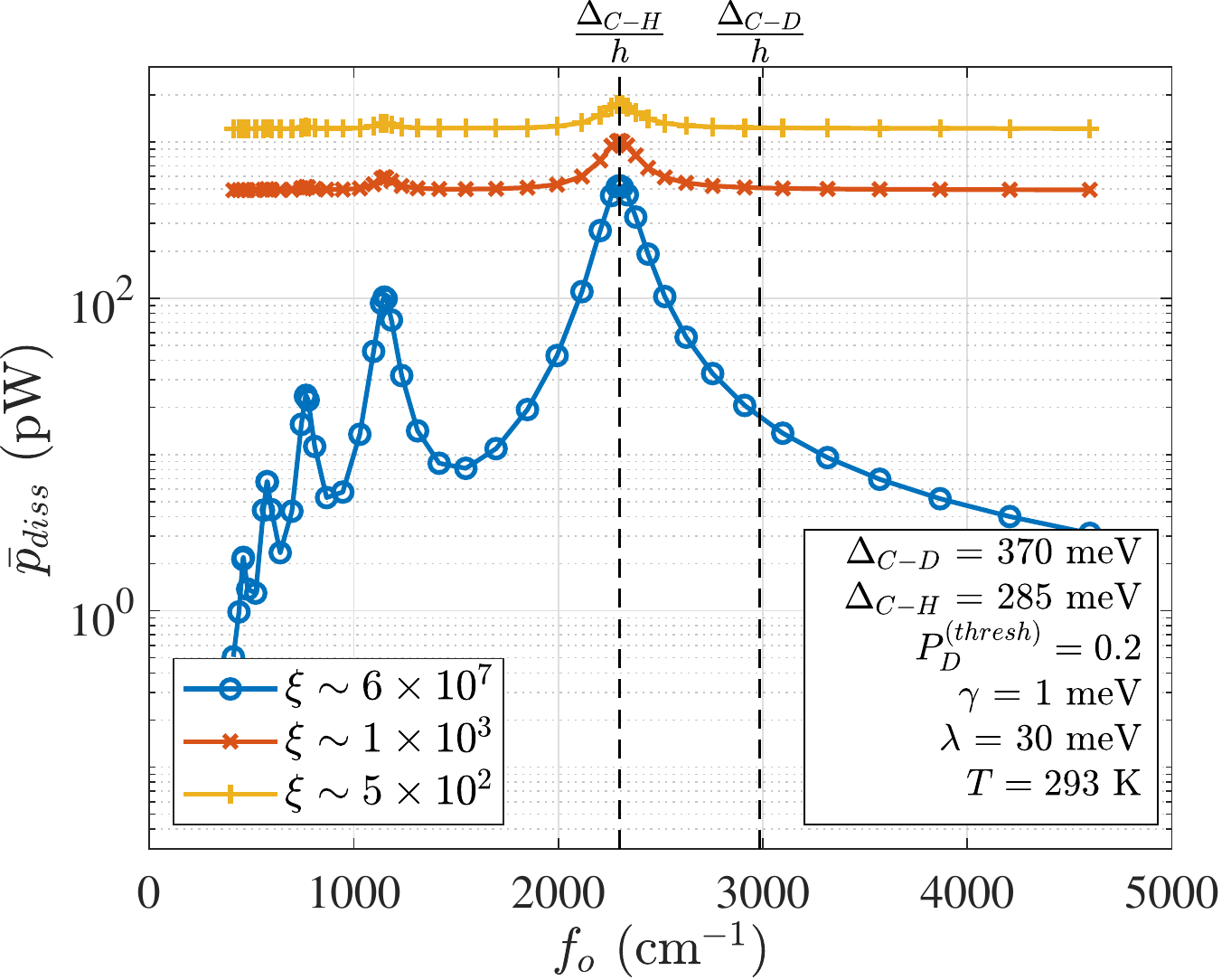}}
\caption{Spectroscopic effects in power dissipation are masked with increasing direct-path dissipation (decreasing $\xi$) for receptors tuned to a specific isotopomer. Subfigure (a) is for a receptor tuned to the C-D stretch frequency, and (b) is for a receptor tuned to the C-H stretch frequency.}
  \label{fig:compare_power_cd&ch}
\end{figure}

% -----------------------------------------------------------------
The degradation of the receptor spectroscopy is also seen in the dissipation plots of Figure \ref{fig:compare_power_cd&ch}. For high $\xi$, the spectroscopic behaviors are readily seen in strong peaks in the dissipation curve $\bar{p} \TextSub{diss} \Argum{f_o}$. As $\xi$ is reduced, spectroscopic behaviors are less pronounced as direct-path dissipation begins to compete with the indirect-path dissipation, enabling higher levels of dissipation in previously-suppressed (non-resonant) frequencies. Eventually, the spectroscopic behavior is completely masked as $\xi$ is further reduced.

% ------------------------------------------------------------ 

% ------------------------------------------------------------ 

The loss of spectroscopic behaviors with reduced $\xi$ can be quantified in the selectivities of the receptors tuned to either the C-D or C-H stretch modes. Selectivity for these receptors is listed in Tables \ref{table:tuned_to_cd} and \ref{table:tuned_to_ch} for the case where $\chi_v = 1/2$. High values of $\xi$ yield the receptor selectivities seen in the indirect-path-dominant regime (see Table \ref{table:selectivity_weak}). When $\xi$ is reduced enough, selectivity is degraded so that isotopomers are no longer distinguishable (selectivity is below $\sim 3$ dB).

% ------------------------------------------------------------ 
\begin{table}[ht!]
\centering
\textbf{Receptor tuned to C-D stretch}  \\[0.5ex] 
\begin{tabular}{||c|c ||} 
\hline
 \textbf{Indirect to direct }  & \textbf{Selectivity}  \\
 \textbf{coupling ratio, $\xi$}&\textbf{(dB)}  \\[0.5ex] 
\hline\hline
$6\times 10^7$ &  16.0 \\
\hline
$2 \times 10^3$  &  3.0\\   
 \hline
$1\times 10^3 $ &  1.5 \\[1ex]       
 \hline 
\end{tabular}
\caption{Strengthening the direct electron-environment dissipation path (reducing $\xi$) degrades the sensitivity of a receptor tuned to the C-D stretch frequency. }
\label{table:tuned_to_cd}
\end{table}
% ------------------------------------------------------------ 
% ------------------------------------------------------------ 
\begin{table}[ht!]
\centering
\textbf{Receptor tuned to C-H stretch}  \\[0.5ex] 
\begin{tabular}{||c|c ||} 
\hline
 \textbf{Indirect to direct }  & \textbf{Selectivity}  \\
 \textbf{coupling ratio, $\xi$}&\textbf{(dB)}  \\[0.5ex] 
\hline\hline
$6\times 10^7$ &  14.8 \\   
\hline
$1\times 10^3$ &  3.0 \\  
 \hline
$ 5\times 10^2$ &  1.5\\[1ex]  
 \hline 
\end{tabular}
\caption{Strengthening the direct electron-environment dissipation path (decreasing $\xi$) degrades the sensitivity of a receptor tuned to the C-H stretch frequency.}
\label{table:tuned_to_ch}
\end{table}
% ------------------------------------------------------------ 

% ------------------------------------------------------------ 

\section{CONCLUSIONS}
A numerical model was presented for olfactory IETS in which the two-state electronic system was coupled to a thermal environment in two ways: (1) directly, and (2) indirectly via a quantum harmonic oscillator modeling the environmentally-damped dominant vibrational mode of the odorant molecule. This model treats explicitly a dominant odorant vibrational mode and provides an method for calculating dissipation from the electron to the environment via both pathways. Resonances were observed in the ET rate for odorant frequencies $\omega_o = \Delta/s \hbar$ when indirect electron-environment coupling is dominant, consistent with the vibrational theory of olfaction. The resonance in the ET process is related to the system's ability to dissipate energy to the environment via the indirect pathway only. In this limit, the model also demonstrates a transition between semi-classical and quantum behavior based on the size of the quantization $\hbar \omega_o$. Odorants with small vibrational quantization $\hbar \omega_o $ exhibit an ET rate behavior in which decreasing temperature reduces the ET rate, consistent with classical and semi-classical ET rate models where thermal excitation provides the means to pass over a potential energy barrier. On the other hand, when $\hbar \omega_o \sim \Delta$, odorant quantization becomes significant, and reducing the temperature of the reservoir has less effect on ET rate because quantum mechanical tunneling becomes a more dominant effect. Also, this model predicts significant ET rates even at $T=0$ K, contrary to Fermi's golden rule. Behavior like this is unique to quantum mechanical models. Spectroscopic behaviors in the ET rate become masked as direct electron-enviornment coupling plays a stronger role in the ET. Nonetheless, the spectroscopic behaviors may be relevant in some biological systems for which the direct dissipative path is negligible.

The vibrational theory of olfaction is an interesting application of quantum mechanics that requires further experimental techniques and  developments for validation. Experimental work will be required to validate this and other models of IETS, and even the vibrational theory of olfaction itself. No experiment has conclusively demonstrated a spectroscopic olfactory mechanism, nor identified the existence of the $D$ and $A$ sites, nor eliminated perireceptor events in distinguishing isotopomers. Further developments in the model could incorporate the inclusion of additional odorant vibrational modes, as well as a treatment of chirality in enantiomers. Models like this may also be used in exploring spectroscopic effects in GCPR more broadly in mammalian nervous systems beyond olfactory receptors.\cite{2014NeuroreceptorActivation}

\section*{ACKNOWLEDGMENT}

The authors thank Craig S.\ Lent of the University of Notre Dame for discussion on the topics of open quantum systems applied to this model.  This work was supported by Baylor University under a new-faculty-startup grant.

%%%%%%%%%%%%%%%%%%%%%%%%%%%%%%%%%%%%%%%%%%%%%%%%%%%%%%%%%%%%%%%%%%%%%%%%%%%%%%%%
  %\addtolength{\textheight}{-12cm}   % This command serves to balance the column lengths
                                  % on the last page of the document manually. It shortens
                                  % the textheight of the last page by a suitable amount.
                                  % This command does not take effect until the next page
                                  % so it should come on the page before the last. Make
                                  % sure that you do not shorten the textheight too much.

\bibliographystyle{IEEEtran}
\bibliography{QuantumOlfactionBibliography}

%%%%%%%%%%%%%%%%%%%%%%%%%%%%%%%%%%%%%%%%%%%%%%%%%%%%%%%%%%%%%%%%%%%%%%%%%%%%%%%%

%%%%%%%%%%%%%%%%%%%%%%%%%%%%%%%%%%%%%%%%%%%%%%%%%%%%%%%%%%%%%%%%%%%%%%%%%%%%%%%%

%%%%%%%%%%%%%%%%%%%%%%%%%%%%%%%%%%%%%%%%%%%%%%%%%%%%%%%%%%%%%%%%%%%%%%%%%%%%%%%%

\end{document}